\documentclass[twocolumn,showpacs,preprintnumbers,amsmath,amssymb,prb,superscriptaddress]{revtex4}

\usepackage{graphicx}
\usepackage{dcolumn}
\usepackage{bm}

\begin{document}

\title{Phase diagram of the Heisenberg spin ladder with ring exchange} 

\author{V. Gritsev} 
\affiliation{D\'epartement de Physique, Universit\'e de Fribourg, 
P\'erolles, CH-1700 Fribourg, Switzerland}

\author{B. Normand}
\affiliation{D\'epartement de Physique, Universit\'e de Fribourg, 
P\'erolles, CH-1700 Fribourg, Switzerland}

\author{D. Baeriswyl}
\affiliation{D\'epartement de Physique, Universit\'e de Fribourg, 
P\'erolles, CH-1700 Fribourg, Switzerland}

\begin{abstract}
We investigate the phase diagram of a generalized spin-1/2 quantum 
antiferromagnet on a ladder with rung, leg, diagonal, and ring-exchange 
interactions. We consider the exactly soluble models associated with the 
problem, obtain the exact ground states which exist for certain parameter 
regimes, and apply a variety of perturbative techniques in the regime 
of strong ring-exchange coupling. By combining these approaches with 
considerations related to the discrete $Z_{4}$ symmetry of the model, 
we present the complete phase diagram.    
\end{abstract}



\maketitle

\section{Introduction}

A number of experiments conducted in recent years suggest that multiple 
spin-exchange interactions have a significant role in the quantitative 
description of the physics of low-dimensional cuprate compounds, and a 
qualitative one in determining the properties of $^3$He adsorbed on 
graphite surfaces. Inelastic neutron scattering measurements of the 
spin-wave spectrum in the two-dimensional (2D) cuprate system 
La$_2$CuO$_4$,\cite{Col} and in the quasi-1D spin-ladder compound 
La$_6$Ca$_8$Cu$_{24}$O$_{41}$,\cite{Mat} as well as two-magnon Raman 
scattering measurements on doped systems related to the latter,\cite{rwk} 
indicate the presence of contributions from a four-spin cyclic exchange 
interaction on the order of 10-20\% of the nearest-neighbor superexchange. 
Measurements of the magnetization and heat capacity of $^3$He films of 
various fillings adsorbed on graphite surfaces\cite{Siq,rcthrbbg} have 
been interpreted\cite{Mis} in terms of cyclic three-, four- and 
higher-spin exchange processes.

The four-spin exchange term upon which we will focus here arises at fourth 
order in a strong-coupling (small $t/U$) perturbative expansion of the 
one-band\cite{Tak} and three-band\cite{MH} Hubbard model in 2D, and has 
been shown in this limit to give the leading correction to the 
nearest-neighbor Heisenberg model. Recent investigations of the influence 
of this interaction on physical properties have employed perturbative 
approaches,\cite{Breh,rmvm} spin-wave analysis,\cite{Chub,rkk} numerical 
exact diagonalization of small clusters,\cite{Mis} exact diagonalization 
in combination with conformal field theory (CFT),\cite{rhn} and the density 
matrix renormalization group (DMRG) technique.\cite{DMRG,dual,Lah,rhqn} 
The majority of these studies is restricted to the regime of weak 
ring-exchange coupling and to systems with only a nearest-neighbor 
Heisenberg superexchange interaction in addition to this term.

Here we consider a generalized model to gain further insight into the 
nature of the phases and phase transitions within this class of system. 
We investigate a spin ladder which includes antiferromagnetic Heisenberg 
leg, rung and diagonal, or cross-plaquette next-neighbor, interactions, 
as well as a ring-exchange term. This system, depicted schematically in 
Fig.~1, represents the minimal model possessing both the possibility of 
a four-spin cyclic exchange interaction and, by virtue of the diagonal
coupling, points in parameter space with non-trivial exact solutions. 

\begin{figure}[t!]
\begin{center}
\includegraphics[width=7.0cm]{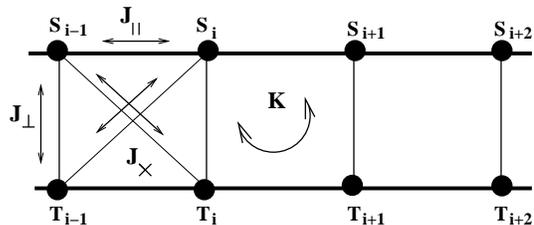}
\caption{Generalized spin-ladder system with rung, leg, and diagonal 
superexchange interactions, $J_{\perp}$, $J_{||}$, and $J_{\times}$, and 
four-spin ring-exchange interaction, $K$. Each site contains a spin $S$ 
= 1/2.\label{ladder}}
\end{center}
\end{figure}

The ring-exchange interaction may be considered to be composed of 
different four-spin coupling terms between the spins of a plaquette, 
subject to the special condition that the effective leg-leg coupling 
is equal to the rung-rung interaction, and equal in magnitude but 
opposite in sign to the diagonal-diagonal coupling. Relaxing this 
constraint on the possible four-spin terms results in a more general 
parameter space with still richer phase behavior, but we are unaware 
of studies of this model for arbitrary values of all coupling parameters. 
In addition to the ring-exchange term, a further possible section of this 
general parameter space is given by the composite-spin representation 
of a $S = 1$ bilinear-biquadratic chain,\cite{rlfs} which corresponds 
to the choice of equal leg-leg and diagonal-diagonal couplings. For this 
specific choice there is an explicit mapping between the spin-1/2 ladder 
model and the spin-1 chain. In general, the different phases of the spin 
model, and the transitions between these, may be distinguished by 
considering the expectation values of nonlocal quantities, such as the 
string order parameter first introduced\cite{rnr} in studies of the spin-1 
chain and the Lieb-Schulz-Mattis twist operator.\cite{rnt} The differing 
effects of the four-spin coupling terms on string order parameters defined 
for the spin-1/2 ladder were analyzed in Ref.~\onlinecite{rfls}. 

\begin{figure*}[t]
\begin{center}
\includegraphics[width=15cm]{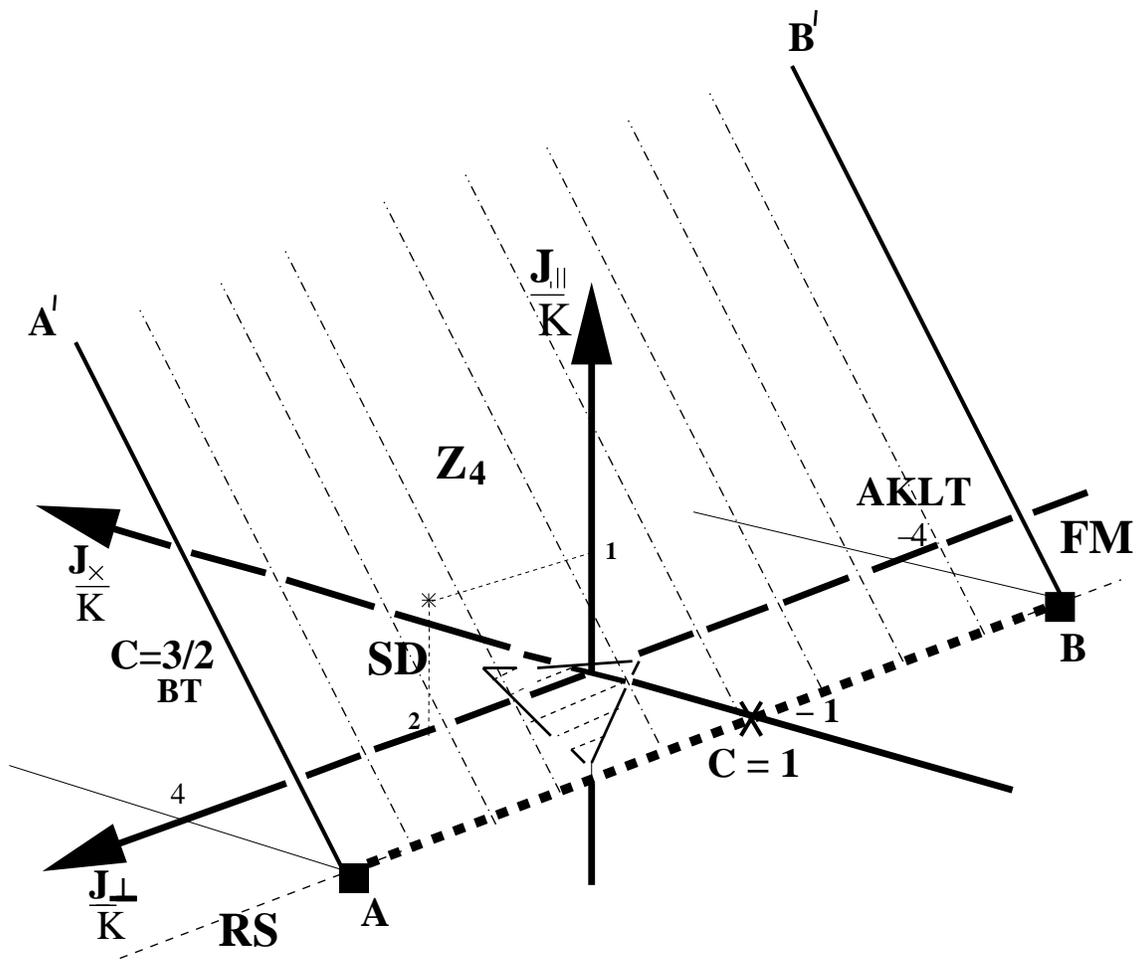}
\caption{Phase diagram of the generalized ladder model of Eq.~(1), 
represented as a function of the coupling ratios $J_{\perp}/K$, $J_{||}/K$, 
and $J_{\times}/K$. The inclined plane indicated by dot-dashed lines is 
invariant with respect to a $Z_{4}$ symmetry, and is given by the equation 
$J_{||}-J_{\times} = K$. The dashed line passing through ${\bf A}$ and 
${\bf B}$ represents an exactly soluble model with central charge $c = 1$ 
between ${\bf A}$ and ${\bf B}$, a ``rung-singlet'' (RS) phase to the left 
of point ${\bf A}$ and a ferromagnetic phase (FM) to the right of point 
${\bf B}$. The full line ${\bf AA'}$ corresponds to a $c = 3/2$ conformal 
field theory and belongs to the Babudjian-Takhtajan (BT) universality class; 
it is the exact transition line between the rung-singlet phase and a 
staggered dimer (SD) phase. The full line ${\bf BB'}$ separates the 
ferromagnetic phase from a spin-gapped region of the type proposed by 
Affleck, Kennedy, Lieb, and Tasaki (AKLT). The shaded triangle lying below 
the $Z_{4}$-symmetric plane represents a critical surface which is also of 
the $c = 3/2$ BT universality class.}
\label{PD}
\end{center}
\end{figure*}

We begin by considering the points within the parameter space for which 
exactly soluble models exist, and then analyze those regimes in the vicinity 
of these points which are accessible by perturbative techniques based on the 
CFT corresponding to each exact solution, {\it 
i.e.}~to the regions where the soluble models are critical. The ladder 
Hamiltonian which we study, and its associated soluble models, are based on 
a plaquette algebra, by which is meant an algebra of operators defined on the 
four-spin plaquette units of the ladder. This algebra satisfies quasi-local 
commutation relations, meaning that operators defined on next- (and 
further-)neighbor plaquettes commute. Using the commutation relations 
of this algebra we have found exact solutions to the Yang-Baxter 
equation.\cite{GB} These solutions provide soluble, $SU(2)$-invariant, 
isotropic spin-ladder Hamiltonians, whose excitation spectrum may be either 
gapless or gapped.\cite{GB} 

Gapless, or critical, solutions arise at points or lines of phase 
transitions, and may be further mapped in the continuum limit to CFTs 
with a central charge $c\geq 1$, or more specifically to Wess-Zumino-Witten 
(WZW) models.\cite{AH} Perturbative approaches constructed around such WZW 
models enable one to access regions of parameter space away from these
second-order phase-transition points. According to the Zamolodchikov 
$c$-theorem,\cite{Zamol} the addition of perturbative conformally 
noninvariant terms represented by relevant and marginal operators 
results in a flow either to another CFT with smaller central charge $c$ 
or to a massive phase. A renormalization-group (RG) procedure constructed 
on this basis of these perturbations may then be used to establish which 
fixed points are stable, and to determine the connectivity, the flow 
structure and thus the physical properties of different regimes of the 
phase diagram. 

An exact description of the gapped phases is possible by using the idea of 
``words'' on the plaquette algebra to construct variational ground states. 
The direct relation of this concept to the matrix-product ansatz\cite{rahy}
allows one to obtain both the exact ground state in suitable parameter 
regimes and the lowest excited states, from which the lines of phase 
transitions may also be deduced. For parameter regimes in the vicinity of 
these exact, gapped solutions, the absence of phase transitions means that 
many qualitative properties of the associated, gapped phases are 
known.

Throughout this study we emphasize the role of different symmetries, 
in particular the $Z_{4}$ symmetry of the model, in obtaining a complete 
understanding of the phase diagram. The relevance of this symmetry to the 
ring-exchange problem was first recognized in Ref.~\onlinecite{dual} in the 
more restricted context of a ``duality'' transformation about a self-dual 
point in the parameter space of the nearest-neighbor ladder model. Here we use 
the full transformation to relate different soluble models, and to develop 
a perturbative bosonization scheme applicable in the regime of strong 
ring-exchange coupling. 

All of these considerations allow us to deduce the complete phase diagram 
for the system in the 3D space of coupling ratios $J_{\perp}/K$, $J_{||}/K$, 
and $J_{\times}/K$, which for the orientation of the reader is presented in 
Fig.~\ref{PD}. The symbols and lines summarized in the caption of 
Fig.~\ref{PD}, and the phases and transitions they represent, are 
explained in the course of the analyses to follow. Fig.~\ref{PD} contains 
as a subset [in the line ($J_{\perp}/K = J_{||}/K$, $J_{\times}=0$)] the 
$K > 0$ part of the phase diagram obtained in the DMRG studies of L\"auchli 
{\it et al.},\cite{Lah} and confirms the presence of all of the phases and 
phase transitions proposed in that work. Placing this subset in the context 
of the full phase diagram permits a significant expansion of our 
understanding of the nature of these phases, and of the first- and 
second-order transitions separating them. Further, it provides certain 
novel, additional phases and transitions inaccessible from the restricted 
parameter space, yielding additional insight into the competition of 
superexchange and ring-exchange interactions. 

In Sec.~II we introduce the Hamiltonian of the model and summarize briefly 
its associated exact solutions, all of which are critical and belong to 
different universality classes. Further exact and variational ground states 
obtained in different regions of parameter space using adapted matrix-product 
wave functions are presented in Sec.~III, accompanied by a discussion of the 
nature of the associated gapped states and their behavior at phase boundaries. 
Sec.~IV contains a complete perturbative analysis around the four available 
CFT solutions, with emphasis on the properties of the strong-coupling 
(large-$K$) regime, which establishes the fixed-point structure of the 
phase diagram. Sec.~V presents a summary and discussion of the results.

\section{Ladder Models}

\subsection{Ring Exchange}

The Hamiltonian of the two-leg ladder (Fig.~1) is
\begin{eqnarray}\label{ladham} 
H & = & J_{||}\sum_{i}({\bf S}_{i} {\bf \cdot S}_{i+1} + {\bf T}_{i}
{\bf \cdot T}_{i+1})\nonumber\\
& & + \, J_{\times} \sum_{i}({\bf S}_{i} {\bf \cdot T}_{i+1} + {\bf T}_{i} 
{\bf \cdot S}_{i+1})\nonumber\\
& & + \, J_{\perp} \sum_{i}{\bf S}_{i} {\bf \cdot T}_{i} \nonumber\\
& & + \, {\textstyle \frac{1}{2}} K \sum_{\Box }(P_{4} + P^{-1}_{4}),
\end{eqnarray}
where in the last (``ring-exchange'') term the four-site permutation 
operator $P_{4}$ exchanges spins in a cyclic manner around each elementary 
plaquette $\Box$ of the ladder, and is given by the equation
\begin{eqnarray}\label{P}
P_{4} + P_{4}^{-1} = 
& & {\bf S}_{i} {\bf \cdot S}_{i+1} + {\bf T}_{i} {\bf \cdot T}_{i+1} 
+ {\bf S}_{i} {\bf \cdot T}_{i} \nonumber\\
& & + \, {\bf S}_{i+1} {\bf \cdot T}_{i+1}+ {\bf S}_{i} {\bf \cdot T}_{i+1} 
+ {\bf T}_{i} {\bf \cdot S}_{i+1} \nonumber\\ & & 
+ \, 4[({\bf S}_{i} {\bf \cdot S}_{i+1})({\bf T}_{i} {\bf \cdot T}_{i+1}) \\
& & + \, ({\bf S}_{i} {\bf \cdot T}_{i})({\bf S}_{i+1} {\bf \cdot T}_{i+1}) 
\nonumber\\ 
& & - \, ({\bf S}_{i} {\bf \cdot T}_{i+1})({\bf S}_{i+1} {\bf \cdot T}_{i})] 
+ {\textstyle \frac{1}{4}}. \nonumber
\end{eqnarray}
We introduce the orthonormal basis
\begin{eqnarray}
|0\rangle = {\textstyle \frac{1}{\sqrt{2}}}(|\uparrow\downarrow \rangle 
- |\downarrow\uparrow\rangle ), \ \ \ \  |1\rangle = |\uparrow\uparrow\rangle, 
\nonumber\\
|2\rangle = {\textstyle \frac{1}{\sqrt{2}}}(|\uparrow\downarrow\rangle 
+ |\downarrow\uparrow\rangle ),\ \ \ \ |3\rangle = |\downarrow\downarrow 
\rangle,
\end{eqnarray}
of singlet and triplet states on each rung, and construct a set of 
corresponding  projection operators which generate the $SU(4)$ algebra
\begin{eqnarray}\label{su4}
X^{\alpha\beta}_i & = &(|\alpha\rangle\langle\beta|)_i, \ \ \ \  
X^{\alpha\beta}_{i} X^{\delta\gamma}_{i} = \delta^{\beta\delta} 
X^{\alpha\gamma}_{i}, \\ \sum_{\alpha}X^{\alpha\alpha}_{i}&=&1, \ \ \ \ 
[X^{\alpha\beta}_{i},X^{\gamma\delta}_{j}] = (\delta^{\beta\gamma} 
X^{\alpha\delta}_{i} - \delta^{\alpha\delta} X^{\gamma\beta}_{j}) \delta_{ij}, 
\nonumber
\end{eqnarray}
where $\alpha,\beta = 0, 1, 2, 3$. Spin operators on both chains of the 
ladder may be expressed in terms of this operator basis.\cite{GB} The ladder 
Hamiltonian (\ref{ladham}) is thus equivalent to a generalized, four-state 
chain with only nearest-neighbor interactions. It can be written as   
\begin{eqnarray}\label{algham}
H & = & \sum_{i} {\textstyle \frac{1}{2}} J_{||} (P^{||}_{i,i+1} - 
E^{-}_{i,i+1})+ {\textstyle \frac{1}{2}} J_{\times} (P^{\times }_{i,i+1} 
- E^{+}_{i,i+1}) \nonumber \\ & & + {\textstyle \frac{1}{2}} K (P_{i,i+1}^{||} 
- E^{+}_{i,i+1}) - ( {\textstyle \frac{1}{2}} J_{\perp} + K)(X^{00}_{i} + 
X_{i+1}^{00}) \nonumber \\ & & +2 K X^{00}_{i} X^{00}_{i+1} + {\textstyle 
\frac{1}{2}} K , 
\end{eqnarray}
where $P^{||}_{i,i+1}$ and $P^{\times }_{i,i+1}$ are respectively permutation 
operators corresponding to ladder legs and plaquette diagonals, and the 
operators $E_{i,i+1}^{\pm}$ are unnormalized projectors on plaquette-singlet 
states. These operators satisfy simple algebraic relations.\cite{cr} Their 
explicit expressions are
\begin{eqnarray}\label{llpo}
P_{i,i+1}^{||} & = & \sum_{\alpha,\beta = 0}^{3} X^{\alpha\beta}_{i} 
X^{\beta\alpha}_{i+1}, \nonumber\\
P_{i,i+1}^{\times} & = & \sum_{\alpha,\beta = 0}^{3}(1 - 2\delta_{\alpha 0}) 
(1 - 2\delta_{\beta 0}) X^{\alpha\beta}_{i} X^{\beta\alpha}_{i+1}, 
\end{eqnarray}
and
\begin{eqnarray}\label{TLoper}
E_{i,i+1}^{\pm} = 2|\psi_{i,i+1}^{\pm}\rangle\langle\psi_{i,i+1}^{\pm}|,
\end{eqnarray}
where the functions $|\psi_{i,i+1}^{\pm}\rangle$ are plaquette-singlet states, 
{\it i.e.} states with total spin per plaquette $S^{2}_{\Box} = ({\bf S}_{1} 
+ {\bf S}_{2} + {\bf S}_{3} + {\bf S}_{4})^{2} = 0$, and may be expressed 
in terms of the rung states $|\alpha\rangle_{i}$ and $|\beta\rangle_{i+1}$ as 
\begin{eqnarray}\label{plaqsing}
|\psi_{i,i+1}^{\pm} \rangle & = & {\textstyle \frac{1}{2}} \{|0\rangle_{i}
|0\rangle_{i+1} \pm |2\rangle_{i}|2\rangle_{i+1} \nonumber \\
& & \mp |1\rangle_{i} |3\rangle_{i+1} \mp |3\rangle_{i}|1\rangle_{i+1} \}.
\end{eqnarray}
There are precisely two possible plaquette-singlet states, corresponding 
respectively to $E_{i,i+1}^{+}$ and $E_{i,i+1}^{-}$. 

\subsection{Integrable Models}

The complete set of exactly soluble isotropic ladder models with short-range 
interactions has been found in Ref.~\onlinecite{GB} using the algebraic 
Bethe-Ansatz method (for a review see Ref.~\onlinecite{Fad}). Not all of 
these models are in the same class as that specified by Eq.~(\ref{ladham}), 
and here we present those cases relevant to a system with ring-exchange 
interactions. All of these are critical, and may be distinguished according 
to their central charge $c$, which characterizes the different universality 
classes close to integrable points.

\subsubsection{\mbox{\boldmath $c = 2$} }

The exactly soluble models of this type have the simple form 
\begin{equation}
H = \sum_{i}(P_{i} - E_{i}).
\label{BR} 
\end{equation}
where $P_{i}$ and $E_{i}$ denote respectively $P_{i,i+1}^{||}$, 
$P_{i,i+1}^{\times}$ and $E_{i,i+1}^{+}$, $E_{i,i+1}^{-}$. 
We thus obtain four different soluble Hamiltonians which correspond to 
the four combinations $\{ P^{||}_{i},E_{i}^{+}\}$, $\{ P^{||}_{i},
E^{-}_{i}\}$, $\{ P^{\times}_{i}, E^{+}_{i}\}$, and $\{ P^{\times}_{i}, 
E^{-}_{i}\}$. Two of these are trivially soluble, because  $\sum_{i} 
(P^{||}_{i}-E^{-}_{i})$ is the Hamiltonian of two decoupled chains and 
$\sum_{i}(P^{\times}_{i}-E^{+}_{i})$ is the same pair of chains 
intertwined by the transformation $S_{2i} \leftrightarrow T_{2i}, 
S_{2i+1} \leftrightarrow S_{2i+1}, T_{2i+1} \leftrightarrow T_{2i+1}$ 
[Eq.~(\ref{algham})]. The Hamiltonian of two decoupled chains is represented 
on the phase diagram of Fig.~\ref{PD} by the point $J_{||}/K =\infty$, 
with all other couplings equal to zero. Nonzero values of these couplings 
generally induce a spin gap, but because of the competing nature of 
different interaction terms there is a possibility of critical behavior 
for some combinations of these couplings. This will be demonstrated 
explicitly in Sec.~IV.  Of the remaining two solutions within the 
general ansatz (\ref{BR}), one is given by the combination $\{ P^{||}_{i},
E_{i}^{+}\}$, which in terms of spin operators is  
\begin{eqnarray}\label{HB}
H_D & =& \sum_{i}{\bf S}_{i} {\bf \!\cdot\! S}_{i+1} 
+ {\bf T}_{i} {\bf \!\cdot\! T}_{i+1}
+{\bf S}_{i} {\bf \!\cdot\! T}_{i+1} 
+ {\bf T}_{i} {\bf \!\cdot\! S}_{i+1}\nonumber\\
& &\!\!\!+4[({\bf S}_{i} {\bf \!\cdot\! S}_{i+1})
({\bf T}_{i} {\bf \!\cdot\! T}_{i+1}) 
-({\bf S}_{i} {\bf \!\cdot\! T}_{i+1})
({\bf S}_{i+1} {\bf \!\cdot\! T}_{i})].\nonumber\\
\end{eqnarray}  
The last, $\{ P^{\times}_{i}, E^{+}_{i}\}$ is obtained by the same 
intertwining transformation from the previous Hamiltonian, and takes the form
\begin{eqnarray}\label{BT}
\!H_{DI}&=&\sum_{i}{\bf S}_{i} {\bf \!\cdot\! T}_{i+1} 
+ {\bf T}_{i} {\bf \!\cdot\! S}_{i+1}
+{\bf S}_{i} {\bf \!\cdot\! S}_{i+1} 
+ {\bf T}_{i} {\bf \!\cdot\! T}_{i+1}\nonumber\\ 
& &\!\!\!\!\!+4[({\bf S}_{i} {\bf \!\cdot\! T}_{i+1})
({\bf T}_{i} {\bf \!\cdot\! S}_{i+1})
- ({\bf S}_{i} {\bf \!\cdot\! S}_{i+1})
({\bf T}_{i} {\bf \!\cdot\! T}_{i+1})].\nonumber\\
\end{eqnarray} 

For all four massless models one obtains two decoupled Bethe-Ansatz 
equations which correspond to the $D_2 = SU(2)\times SU(2)$ algebra. 
The energy of the Hamiltonian is the sum of two expressions  for each 
$SU(2)$ component,\cite{martins} and thus the Hamiltonian (\ref{HB}) 
is equivalent to two decoupled chains. This model is critical (no spin 
gap), with a conformal charge $c = 2$. Eq.~(\ref{HB}) constitutes a part 
of the plaquette term, Eq.~(\ref{P}), while we will demonstrate explicitly 
below that the remaining contributions to $P_4 + P_{4}^{-1}$ induce a spin 
gap, and therefore that the origin of coordinates in the phase diagram of 
Fig.~\ref{PD} represents a gapped system. However, as shown in Sec.~IV, 
the competition between different interaction terms of the Hamiltonian 
(\ref{ladham}) drives the system to a second-order phase transition 
described by a CFT with $c = 3/2$. The approximate form of this critical 
surface, represented in Fig.~\ref{PD} by the shaded triangle, is obtained 
in Sec.~IV.     

Exact solutions in this class then provide two massless models, one in 
the limit of weak $K$, and one in the strong-$K$ limit, both of which 
correspond in the continuum limit to $c = 2$ WZW models. The four solutions 
of the (\ref{BR}) are related by two transformations, one 
of which is the intertwining transformation shown above. The other is less 
transparent in nature, and is found by first noting that the two unnormalized 
projectors $E_{i}^{+}$ and $E^{-}_{i}$ are related by the $X$-operator 
transformation
\begin{equation}
X^{0a}_k\rightarrow -i X_{k}^{0a} ,\ X^{a0}_k\rightarrow i X_{k}^{a0}, 
\end{equation}
for $a=1,2,3$ and $k=1,\dots,N$, where $N$ is the total number of sites. 
This is a unitary transformation generated by the operator
\begin{eqnarray}\label{Z4}
U(\pi/2)= \exp[-i\frac{\pi}{2}\sum_{k=1}^N(X^{00}_k )],
\end{eqnarray}
which has the property that
\begin{eqnarray}
 H_{D}= U H_{2c}U^{\dag},
\label{eu}
\end{eqnarray}
where $H_{2c}$ is the Hamiltonian of two uncoupled spin-1/2 Heisenberg 
chains. The nature of this transformation may be understood from the 
observation that it maps one of the plaquette-singlet state defined in 
Eq.~(\ref{plaqsing}) into the other, and thus that it transforms the 
projectors (\ref{TLoper}) according to $E^{+}_{i} \leftrightarrow E^{-}_{i}$.
This transformation has the additional property that $[U(\pi/2)]^4 = 1$, 
and therefore $U$ is one of the generators of the $Z_{4}$ transformation 
associated with the center of $SU(4)$. In the next section we will 
demonstrate that  this symmetry is essential for a complete understanding 
of the phase diagram. 
 
\subsubsection{\mbox{\boldmath $c = 1$}}

A second exact solution\cite{GB} has the same eigenspectrum, but not the 
same degeneracies, as the spin-1/2 Heisenberg chain. The corresponding spin 
Hamiltonian, 
\begin{eqnarray}\label{1/2}
H_{1/2} \!\! & = & \!\! - K \sum_{i}( {\bf S}_{i} {\bf \cdot T}_{i+1} 
+ {\bf T}_{i} {\bf \cdot S}_{i+1}) \\
& & + {\textstyle \frac{1}{2}} K\sum_{\Box}(P_{4} + P^{-1}_{4})
 + J_{\perp}/2\sum_{i}({\bf S}_{i} {\bf \cdot T}_{i}),\nonumber
\end{eqnarray}
commutes with the generators $\sum_{i} X_i^{ab}$ for any $a,b=1,2,3$, 
which form an $SU(3)$ subalgebra within the $SU(4)$ algebra generated by 
the $X$-operators. In view of completeness relation in Eq.~(\ref{su4}), 
$H_{1/2}$ commutes with the total number operator $\sum_{i} X^{00}_{i}$ 
for singlets, and any multiple of this term  may be added to the 
Hamiltonian without spoiling the integrability.  The full symmetry of 
the model is thus $SU(3) \times U(1) = U(3)$. In the framework of the 
equivalent Heisenberg chain, inspired by the Bethe-Ansatz solution, the 
model may be considered as a chain of effective $SU(2)$ ``spins'' ${\bf 
L}_i$ with 
\begin{equation}
L_i^z \equiv {\textstyle \frac{1}{2}} - X_i^{00} = {\textstyle 
\frac{1}{4}} - {\bf S}_{i} {\bf \cdot T}_{i}. 
\label{elz}
\end{equation}
The term $J_{\perp}/2K \sum_{i} {\bf S}_{i} {\bf \cdot T}_{i}$ in $H_{1/2}$ 
therefore corresponds to a coupling with a magnetic field $h \equiv 
J_{\perp}/2K$.\cite{AFW} It is well known that the spin-1/2 chain in a 
magnetic field develops an incommensurate critical phase for $|h| \leq 2$ 
and has a massive phase for $|h| > 2$ (see for example Ref.~\onlinecite{BF}).

One observes from Eq.~(\ref{1/2}) that this integrable model corresponds to 
the line in parameter space $J_{||} = 0$, $J_{\times} = - K$, for arbitrary 
$J_{\perp}$. When  $(|J_{\perp}/K|) \leq 4$ the model is in the critical, 
incommensurate phase, whereas for $J_{\perp}/K > 4$ the model has a gapped, 
rung-singlet phase, and in the region  $J_{\perp}/K < -4$ it has a gapped, 
rung-triplet ground state. The nature of the gapped phases and the physical 
properties of the incommensurate phase will be discussed in more detail 
in Sec.~III. In the critical region the model is described by a CFT with 
central charge $c = 1$.\cite{Alc} However, the presence of conserved 
charges generating the $U(3)$ symmetry has the consequence\cite{Alc} that 
the model possesses additional zero modes, and therefore while the 
conformal dimensions are those of the $c = 1$ theory for the Coulomb 
gas, the degeneracies are altered accordingly.  The critical region 
is represented on the phase diagram of Fig.~\ref{PD} by the thick, dashed 
line ${\bf AB}$, while the rung-singlet and ferromagnetic rung-triplet 
regions are denoted respectively by RS and FM. In Sec.~IV 
(Fig.~\ref{section}) we will demonstrate that there exists a finite 
critical region in the vicinity of the critical line. 

\subsubsection{\mbox{\boldmath $c = 3$}}

A further exact solution exists which corresponds simply to local 
Hamiltonians proportional to permutation operators $P^{||}_{i}$ and 
$P^{\times}_{i}$. The Hamiltonian corresponding to $P^{||}_{i}$,
\begin{eqnarray}\label{SO}
H_{||}\!&\! =\! &\!\sum_{i}{\bf S}_{i} {\bf \cdot S}_{i+1} 
+{\bf T}_{i} {\bf \cdot T}_{i+1}+ 4 ({\bf S}_{i} {\bf \cdot S}_{i+1}) 
({\bf T}_{i} {\bf \cdot T}_{i+1})\nonumber\\
\end{eqnarray}
is invariant with respect to the full $SU(4)$ group and describes a soluble 
``spin-orbital"  model.\cite{KK} In the continuum limit, Eq.~(\ref{SO}) 
corresponds to an $SU(4)$ WZW model at level $k = 1$,\cite{AGLN,IQA} and 
therefore has central charge $c = 3$. This Hamiltonian commutes with the 
operator $\sum_{i}({\bf S}_{i} {\bf \cdot T}_{i})$, and thus is integrable 
for arbitrary values of $J_{\perp}$. One may also add a rung-rung coupling 
interaction to obtain the Hamiltonian
\begin{eqnarray}\label{w}
H_{||}' \!&\! = \!&\! H_{||} + J_{\perp} \sum_{i} {\bf S}_{i} {\bf \cdot 
T}_{i} + 2 K \sum_{i} ({\bf S}_{i} {\bf \cdot T}_{i}) ({\bf S}_{i+1} {\bf 
\cdot T}_{i+1}).\nonumber\\
\end{eqnarray}
which was considered in Ref.~\onlinecite{Wang} for arbitrary $J_{\perp}$ 
and for two values of $K$ ($K = 0$ and $K = -1$). It was shown that for 
both values of $K$, three phases appear as a function of $J_{\perp}$, and 
are separated by two quantum critical points, $J_{+}^{c}$ and $J_{-}^{c}$. 
For $K = 0$ and $J_{\perp} > J_{+}^{c} = 2$ the model is in the 
rung-dimerized phase, for $J^{c}_{-} < J < J^{c}_{+}$ there is a gapless 
phase with $c = 3$, and for $J < J^{c}_{-} \approx -1.79$ there is a gapless 
phase with $c=2$. For $K = -1$ the critical points occur at\cite{Wang} 
$J_{+}^{c} = 1/2$ and $J_{-}^{c} \approx -1.29$.  Quite generally, the 
extent of the $c = 3$ critical region is expected to be strongly reduced 
by a negative rung-rung coupling, while it will expand for positive $K$. 
The critical behavior of $H_{||}'$ is thus expected to persist for finite 
regions of parameter space, and the position of the critical points to 
depend strongly on the rung-rung coupling. We emphasize that the model of 
Eq.~(\ref{w}) does not appear in the phase diagram of the ladder model 
under consideration (\ref{ladham}) but we have introduced its properties 
here for use in Sec.~IV as one appropriate basis model for a perturbative 
expansion. Similar considerations apply to $H_{\times}$, which is obtained 
from $H_{||}$ by using the intertwining transformation of Sec.~IIB.1.

\begin{figure}[t!]
\begin{center}
\includegraphics[width=7cm]{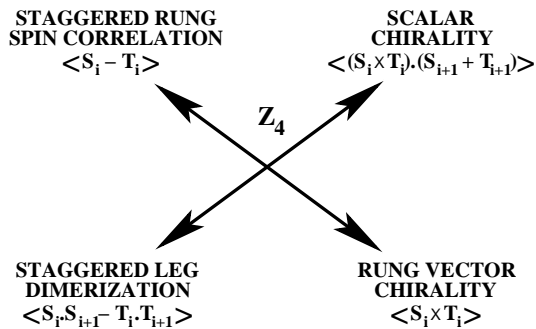}
\caption{Order parameters of the different phases related by the $Z_{4}$ 
transformation within the plane $J_{||} - J_{\times} = K$.\label{duality}}
\end{center}
\end{figure}

\section{Exact ground states}

\subsection{\mbox{\boldmath $Z_4$} Plane}

In this section we apply certain discrete symmetry considerations to obtain 
further insight into the structure of the phase diagram. We employ the 
matrix-product ansatz to find exact ground states, and also the boundaries 
between these, even in the region of parameter space where the Hamiltonian 
(\ref{ladham}) is not exactly integrable.

We begin by noting that for arbitrary $J_{\perp}$, when
\begin{eqnarray}\label{plane}
J_{||} - J_{\times} = K
\end{eqnarray}
the Hamiltonian (\ref{ladham}) is invariant with respect to the $Z_{4}$ 
transformation (\ref{Z4}) which generates the mapping $E_{i}^{\pm} 
\rightarrow E^{\mp}_{i}$. The condition (\ref{plane}) defines a plane 
in the 3D space of the phase diagram, marked in Fig.~\ref{PD} by parallel 
dot-dashed lines. The Hamiltonian of the system in the $Z_{4}$ symmetric 
plane commutes with the singlet total number operator $\sum_{i} X_{i}^{00}$.  

This $Z_{4}$ transformation is precisely the one noted in 
Refs.~\onlinecite{dual} and~\onlinecite{rmhnh}, in which it was referred 
to as a duality transformation. It is a canonical transformation for the 
lowest ($s$=1/2 $\times$ $s$=1/2 ) representation of $SU(2)\times SU(2)$, 
and conserves the values of the Casimir operators in this representation. 
In terms of the original spin variables it takes the form
\begin{eqnarray}\label{dual}
{\bf\tilde{S}}_{i} & = & {\textstyle \frac{1}{2}}({\bf S}_{i} + {\bf T}_{i})
 - {\bf S}_{i} \! \times \! {\bf T}_{i}  , \nonumber\\
{\bf\tilde{T}}_{i} & = & {\textstyle \frac{1}{2}}({\bf S}_{i} + {\bf T}_{i})
 + {\bf S}_{i} \! \times \! {\bf T}_{i}. 
\end{eqnarray}
When expressed using the variables ${\bf\tilde{S}}_{i}, {\bf\tilde{T}_{i}}$, 
the Hamiltonian $H_{D}$ (\ref{HB}) is exactly that of two decoupled spin 
chains. It is clear from the definition (\ref{P}) of the plaquette term 
that $H_{D}$ constitutes a part of the operator $P_{4}+P_{4}^{-1}$, while 
the remaining contributions to this term couple the two effective spin-1/2 
chains to produce a system with a finite spin gap.

\begin{figure}[t!]
\begin{center}
\includegraphics[width =7cm]{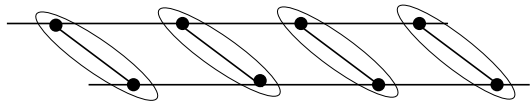}
\caption{Representation of the rung-singlet (RS) phase. The ellipses denote 
a singlet state of the spins on each rung.\label{frs}}
\end{center}
\end{figure}

\begin{figure}[b!]
\begin{center}
\includegraphics[width =7cm]{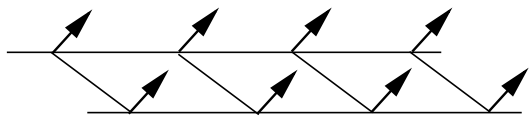}
\caption{Typical ferromagnetic (FM) ground state for large, negative 
$J_{\perp}$.\label{ffm}}
\end{center}
\end{figure}

In the $Z_4$ plane this transformation acts as a symmetry between different 
order parameters. It maps the order parameter $\langle {\bf 
S}_{i} - {\bf T}_{i} \rangle$ for antiferromagnetic rung spin correlations 
into the vector-chirality order parameter $\langle {\bf S}_{i} \! \times \! 
{\bf T}_{i} \rangle \equiv \langle {\bf\tilde{S}}_{i} - {\bf\tilde{T}}_{i} 
\rangle$, while the leg dimer order parameter $\langle {\bf S}_{i} {\bf \cdot 
S}_{i+1} - {\bf T}_{i} {\bf \cdot T}_{i+1} \rangle$ is mapped into the order 
parameter $\langle ({\bf S}_{i}+{\bf T}_{i})({\bf \cdot S}_{i+1} \! \times \! 
{\bf T}_{i+1}) + ({\bf S}_{i} \! \times \! {\bf T}_{i}) ({\bf \cdot S}_{i+1} + 
{\bf T}_{i+1}) \rangle$ for scalar chirality (Fig.~\ref{duality}). Phases 
characterized by order parameters related under the $Z_{4}$ transformation 
within each pair are therefore symmetric within the plane defined by 
Eq.~(\ref{plane}). The point $J_\perp = J_{||} = K$ considered in the 
DMRG studies of Refs.~\onlinecite{dual} and \onlinecite{Lah} lies in 
this plane (the star in Fig.~\ref{PD}), giving the properties of duality 
observed in the results of both analyses. 

\begin{figure}[t!] 
\begin{center}
\includegraphics[width =7cm]{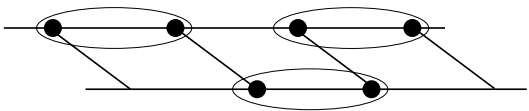}
\caption{Staggered leg-dimerized state (SD).\label{fsd}}
\end{center}
\end{figure}

From this consideration we may deduce the nature of a candidate ground 
state as a product of linear superpositions of the two plaquette-singlet 
states. The complete set of plaquette states with fixed angular momentum 
contains one quintuplet $(j = 2)$, three triplets $(j = 1)$ and two singlets 
$(j = 0)$. The two plaquette-singlet states are those constructed above, the 
states with $j=1$ are created by applying a linear combination of operators 
$P_{i,i+1}^{||}$, $P_{i,i+1}^{\times}$, and $X_{i}^{00}X_{i+1}^{00}$, 
and the state with $j = 2$ is created by the action of a combination of 
the operators  $P_{i,i+1}^{||}$, $P_{i,i+1}^{\times}$, and $E_{i,i+1}^{\pm}$. 
The expressions for these states allow their identification with 
matrix-product-ansatz states, while the explicit form of the Hamiltonian 
(\ref{algham}) corresponds to the operator basis for matrix-product-ansatz 
Hamiltonians.\cite{MPA} One observes that in the $Z_{4}$-symmetric plane, 
$J_{||}-J_{\times} = K$, the condition $(J_{\perp}/K) > 4$ defines a region 
in the phase diagram (Fig.~\ref{PD}) with an exact rung-singlet (RS) ground 
state, 
\begin{eqnarray}\label{sing}
|\psi_{RS}\rangle = |0\rangle_{1} |0\rangle_{2} \dots |0\rangle_{N-1} 
|0\rangle_{N},
\end{eqnarray}
which is represented schematically in Fig.~\ref{frs}. On the line 
$J_{\times} = - K$, $J_{||}=0$, which lies in the $Z_{4}$-symmetric plane, 
this statement is supported by the exact solution of the Hamiltonian 
$H_{1/2}$ (\ref{1/2}) of the previous section. From the results of 
Ref.~\onlinecite{MPA}, the line $J_{\perp}/K = 4$ in the $Z_{4}$-symmetric 
plane is a line of second-order phase transitions into a spontaneously 
dimerized phase of staggered leg dimerization (SD). 

\begin{figure}[b!]
\begin{center}
\includegraphics[width=7cm]{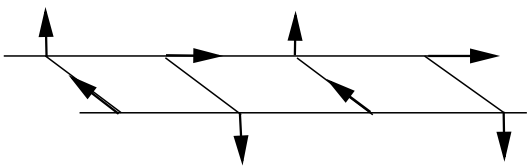}
\caption{Representation of a spin configuration with finite scalar-chirality 
(SC) correlation function.\label{fsc}}
\end{center}
\end{figure}

Similarly, the condition $J_{\perp}/K < - 4$ defines an exact ferromagnetic 
rung ground state (FM),
\begin{eqnarray}\label{trip}
|\psi_{FM}\rangle = X^{a0}_{1}X^{a0}_{2}...X^{a0}_{N-1} X^{a0}_{N}|0\rangle ,
\end{eqnarray}
in which $|0 \rangle$ denotes the global state of singlets on every rung.
This state is represented in Fig.~\ref{ffm}.
The transition on the line $J_{\perp}/K = - 4$ is of first order. Both 
transition lines are determined exactly from the points where the dispersion 
relations of the elementary excitations become massless. In the rung-singlet 
phase this excitation is a propagating rung triplet, while in the 
ferromagnetic phase it is a rung singlet. The second-order transition 
from the rung singlet phase to the staggered dimer phase (Fig.~\ref{fsd}) 
lies in the universality class of the spin-1 bilinear-biquadratic chain, 
which in the continuum limit  is described by a $c = 3/2$ CFT,\cite{AH} 
and is a transition of the Babudjian-Takhtajan type. This transition is 
characterized by a spontaneous breaking of discrete $Z_{2}$ symmetry, 
which in the ladder is related to translation by one a lattice unit.

\begin{figure}[t!]
\begin{center}
\includegraphics[width=7cm]{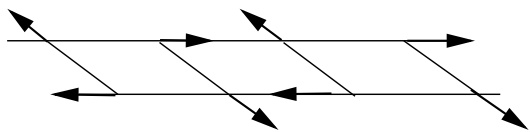}
\caption{Schematic representation of a static vector-chirality 
configuration (VC).\label{fvc}}
\end{center}
\end{figure}

From the $Z_{4}$ duality of the system we deduce that at large $K$ (near the 
origin of Fig.~\ref{PD}) there is another second-order phase transition of 
the same universality class from a scalar-chirality phase (SC) 
(Fig.~\ref{fsc}), which is dual to the staggered dimer phase, to a 
vector-chirality phase (VC) (Fig.~\ref{fvc}) which is dual to the 
rung-singlet phase. In the next section we provide further arguments in 
support of this statement. Figs.~\ref{fsc} and \ref{fvc} show respectively 
typical static configurations of spins in scalar- and vector-chirality 
phases; these should be understood only as indicating the preferred 
instantaneous spin configurations, which in fact fluctuate rapidly, such 
that the phases exist only in the sense of finite average values of the 
corresponding spin correlation functions. All four phases have only 
short-ranged correlations in the chain direction. These states illustrate 
the increasing dominance of the $K$ term, which favors configurations in 
which all spins on a plaquette are mutually perpendicular to maximize their 
solid angle,\cite{Chub} in the competition with nearest-neighbor exchange 
terms which favor antiparallel spins. 

\begin{figure}[b!]
\begin{center}
\includegraphics[width=7cm]{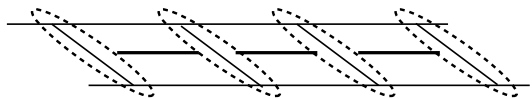}
\caption{AKLT-type ground state. The dashed ellipses represent effective 
spin-1 variables formed by triplet states on the ladder rungs.\label{faklt}}
\end{center}
\end{figure}

Finally, the first-order phase transition from the ferromagnetic phase takes 
the system to a type of Affleck-Kennedy-Lieb-Tasaki (AKLT) state\cite{AKLT} 
formed by effective spin-1 variables represented by the rung triplets 
(Fig.~\ref{faklt}). It differs from the ferromagnetic phase in that 
coherence between the individual rung triplets is not established until 
the transition at $J_\perp/K = - 4$. This state is obtained directly from 
the vector-chirality phase in a first-order transition at small, negative 
$J_\perp$. We stress that the letters indicating the locations of these 
phases in Fig.~\ref{PD} should be understood to refer only to the plane 
of exact $Z_4$ symmetry, including the rung-singlet and ferromagnetic 
phases which lie outside the region marked by the dot-dashed 
lines. However, because there are no phase transitions other than those 
present on the diagram, the gapped phases obtained for model parameters 
outside the plane are connected continuously to those of the exactly known 
states, and their physical properties evolve continuously as the values of 
the interactions are further changed away from the $Z_4$ plane.

\subsection{Incommensurate Line}

With these results, a heuristic understanding of the origin of the 
incommensurate phase appearing on the critical line {\bf AB} in Fig.~\ref{PD} 
(Sec.~II) may be obtained by considering the role of $J_\perp$ as an 
effective magnetic field. We stress first that the incommensurate phase 
arises with no breaking of $SU(2)$ symmetry in the space of the spin 
variables ${\bf S}_i$ and ${\bf T}_i$. The variable conjugate to the 
effective field, $\sum_i {\bf S}_i {\bf \cdot T}_i$, corresponds 
[Eq.~(\ref{elz})] to an average singlet density which 
varies from 0 at the boundary to the ferromagnetic phase ($J_\perp 
\rightarrow - 4 K$) to 1 at the rung-singlet phase boundary ($J_\perp 
\rightarrow 4 K$), in exact analogy to $\sum_i S_i^z$ for the spin-chain 
problem. For the ladder, $\sum_i {\bf S}_i {\bf \cdot T}_i$ also expresses 
the spin correlation on each rung, which may be characterized by defining 
a variable $k_\perp = \cos^{-1} (\langle {\bf S}_i {\bf \cdot T}_i 
\rangle / |{\bf S}_i| |{\bf T}_i|)$. This effective wave vector across the 
ladder varies continuously from 0 to $\pi$ over the range $- 4 K \le J_\perp 
\le 4 K$, and takes the value $k_\perp = \pi/2$ at $J_\perp = 0$, where the 
ring-exchange term, favoring locally perpendicular spins, is dominant. 

The incommensurate state in the ladder direction is crucially dependent 
on the parameters of the $c = 1$ line, notably the special role of the 
$J_{\times}$ term. Only for $J_{\times} = - K$ is the spectrum massless, 
with quasi-long-ranged correlation functions classifiable by a ladder 
wave vector $q$ which reflect the competition between nearest-neighbor 
exchange and ring exchange, again in exact analogy with the spin-1/2 
Heisenberg chain in a magnetic field.\cite{rmtbb} The nature of the 
incommensurate correlations arising from the excitation spectrum is 
elucidated by considering the effective spin variables ${\bf L}_i$. The 
analogous ``spin'' correlation function is given by 
\begin{eqnarray}
\langle {\bf L}_i {\cdot \bf L}_j \rangle & = & \langle X_i^{0a} X_j^{a0} 
+ X_i^{a0} X_j^{0a} \rangle \label{elx} \nonumber \\ & & + 2 \;\! \langle 
({\textstyle \frac{1}{2}} - X_i^{00})({\textstyle \frac{1}{2}} - X_j^{00}) 
\rangle,
\end{eqnarray}
where the first line corresponds to terms of the form $L_i^+ L_j^- + 
L_i^- L_j^+$ and the second to $L_i^z L_j^z$. In terms of the original 
spin variables one obtains 
\begin{eqnarray}
\langle {\bf L}_i {\cdot \bf L}_j \rangle & = & {\textstyle \frac{1}{2}} 
\langle ({\bf S} - {\bf T})_i ({\bf S} - {\bf T})_j \rangle + 2 \langle 
({\bf S} \times {\bf T})_i ({\bf S} \times {\bf T})_j \rangle \label{els} 
\nonumber \\ & & + {\textstyle \frac{1}{2}} \langle ({\textstyle \frac{1}{2}} 
+ 2 {\bf S}_i {\cdot \bf T}_i) ({\textstyle \frac{1}{2}} + 2 {\bf S}_j 
{\cdot \bf T}_j) \rangle . 
\end{eqnarray}
Thus the dynamical correlation function exhibiting a peak at the 
incommensurate wave vector $q$ is in fact the sum of the staggered rung, 
vector chirality, and singlet density correlation functions, although from 
the massless nature of the excitations at $q$ one may expect peaks in each 
function individually. Again the terms in the first line correspond to 
$L_i^+ L_j^- + L_i^- L_j^+$, and are symmetrical under the $Z_4$ 
transformation (Fig.~\ref{duality}) in the plane: their peaks appear at 
the same wave vector for all values of $J_\perp$, but their intensities 
may differ. The effective $L_i^z L_j^z$ terms (second line) are not 
symmetrical with the others for all coupings $J_\perp \ne 0$ (finite 
effective field), but as in the spin chain may be expected to show the 
same continuous evolution of the incommensurate peak position from $q$ = 
0 at $J_\perp = - 4 K$ to $q = \pi$ at $J_\perp = 0$, and back to $q$ = 0 
at $J_\perp = 4 K$. We stress again that this behavior is a specific 
property of the incommensurate line and cannot in general be expected to 
be clearly visible in other parts of the phase diagram not in its immediate 
vicinity, including the regions corresponding to DMRG analyses performed to 
date. However, in Sec.~IV we isolate a region in which the incommensurate 
properties (albeit for a gapped excitation spectrum) may persist over a 
significant range of parameters. 

\section{Renormalization-group analysis}

In this section we study the perturbations around different CFT solutions, 
which either appear in the phase diagram of the model (\ref{ladham}) or are 
closely related to it. We adopt a variety of techniques to address the nature 
of the ground states in regions away from these soluble points, and to 
establish the fixed-point structure of the phase diagram. Because all of 
the transition points and lines are known, the relevance of the operators 
within Eq.~(\ref{ladham}) as perturbations of the exact solutions determines 
the flow under renormalization and thus the dominant physical properties of 
the intermediate regions. In each subsection and in the Appendix, known 
results are summarized briefly while those which are new in the current 
context are presented in detail.

We have found four models suitable for this type of analysis. From 
the exact solutions of Sec.~II, there are two CFTs with $c = 2$ which 
correspond to two decoupled chains described by the soluble Hamiltonian of 
Eq.~(\ref{HB}); one solution is relevant for weak $K$, and the other for 
strong $K$. Here we concentrate primarily on the strong-coupling regime, 
{\it i.e.}~large $K$, for which significantly less is known: the 
fermionized version of the limiting model of two effectively decoupled 
spin chains is analyzed in subsection A by a conventional RG procedure. We 
have argued from the presence of the $Z_4$ symmetry (Secs.~II, III) that 
in the strong-coupling limit there is a second-order phase transition with 
$c = 3/2$, represented by the shaded triangle on the phase diagram of 
Fig.~\ref{PD}. The Zamolodchikov $c$-theorem\cite{Zamol} then demands 
that weak- and strong-coupling limits are disjoint in RG sense, 
{\it i.e.}~that there is no continuous flow from weak to strong coupling. 
To describe the intermediate regime between the two second-order phase 
transitions with $c = 3/2$ we therefore use in subsection B another CFT 
with $c = 3$, which is described by the model (\ref{w}). By perturbative 
analysis of a fermionized model we find that this intermediate region is 
gapped; an alternative perturbative treatment of the $SU(4)$ model 
(\ref{SO}) has recently provided similar results.\cite{rmhnh} 
Finally, the same approach may also be applied to the $c = 1$ CFT which 
corresponds to the solution (\ref{1/2}). In the vicinity of the line 
${\bf AB}$ on the phase diagram of Fig.~\ref{PD}, we employ rather general 
arguments in subsection C to reveal the presence of a massless region in 
one plane of the phase diagram. The CFTs with $c = 3/2$ thus represent 
unstable points in parameter space and must be accompanied by a flow 
towards stable fixed points. Based on the arguments of Ref.~\onlinecite{AH} 
the stable theory should be a $SU(2)_{k = 1}$ CFT with $c = 1$, the natural 
candidate for which is the CFT corresponding to the model of Eq.~(\ref{1/2}). 

\subsection{\mbox{\boldmath $c = 2$} CFTs}

A weak-coupling bosonization analysis is appropriate for the limit of two 
quasi-decoupled chains when $J_{||}\gg K$ and  $J_{||}\gg J_{\perp}$. The 
ladder system with biquadratic exchange has been shown to undergo a 
second-order phase transition at which the behavior of the massless modes 
is governed by a CFT with central charge $c = 3/2$.\cite{NT} The consideration 
of a ring-exchange term is technically identical (see Ref.~\onlinecite{rmvm}), 
and the results are found to be in good agreement with those from other 
approaches.

When performing a perturbative analysis around CFTs, all contributions to 
the spin Hamiltonian should be classified according to the scaling dimensions 
of the operator content of the corresponding CFT. In the case of perturbations 
around the limit of two decoupled chains, in both weak- and strong-$K$ limits, 
four-spin  interaction terms and interleg couplings are expressed in terms 
of two fundamental $c = 1$ WZW fields, $g_{a}$ and $g_{a}'$ ($a$=0,1,2,3), 
with conformal dimension $({\textstyle \frac{1}{4}},{\textstyle \frac{1}{4}})$,
and of Kac-Moody currents ${\bf J}$ and ${\bf\bar{J}}$ with dimensions $(1,0)$ 
and $(0,1)$. This theory can be further expressed in terms of four different 
Ising models,\cite{Sen,NT} {\it i.e.}~of order-disorder fields, energy 
operators and Majorana fermions. The representation of these fields by 
four Ising models is summarized in the Appendix. The connection to the 
spin variables is given by     
\begin{eqnarray}\label{bosspinvar}
{\bf\tilde{S}}_{i}&\rightarrow & a{\bf\tilde{S}}(x) \ \nonumber\\
{\bf\tilde{S}}(x)&=&{\bf J}+{\bf\bar{J}}+(-1)^{x/a} \Theta \mbox{Tr}[g(x) 
{\bf\sigma }]\\
{\bf\tilde{T}}(x)&=&{\bf J}'+{\bf\bar{J}}'+(-1)^{x/a}\Theta \mbox{Tr}[g'(x) 
{\bf\sigma }].\nonumber
\end{eqnarray}
We note that in Eq.~(\ref{bosspinvar}) the uniform and staggered parts of 
the spin-density operators have different conformal dimensions, and that 
$\Theta$ is a nonuniversal normalization constant. 

The expression in terms of Ising-model fermions for the general ladder 
Hamiltonian, which includes arbitrary leg-leg, diagonal-diagonal and 
rung-rung couplings, is derived in the Appendix, and is valid around 
the limits of two decoupled chains. There are two such limits, the weak-$K$
 regime which corresponds to the two initial chains, and the strong-$K$ 
regime which corresponds to two $Z_{4}$-rotated chains (\ref{eu}), 
resulting in the Hamiltonian (\ref{HB}). The expressions of the Appendix 
are valid for both cases. 

The resulting continuum-limit Hamiltonian for the system in weak- and 
strong-coupling regimes may be expressed in terms of four Majorana 
fermions with different singlet and triplet masses,\cite{NT}
\begin{eqnarray}\label{contham}
H & = &  {\textstyle \frac{-i}{2}}\int dx [v_{s}(\psi^{0} \partial_{x} 
\psi^{0} - \bar{\psi}^{0} \partial_{x} \bar{\psi}^{0}) + m_{s}\psi^{0} 
\bar{\psi}^{0} \nonumber \\ & & + \sum_{a=1,2,3}v_{t} 
(\psi^{a} \partial_{x} \psi^{a} - \bar{\psi}^{a} \partial_{x} 
\bar{\psi}^{a}) + m_{t} \psi^{a} \bar{\psi}^{a}] \nonumber \\
& & + H_{\rm{marg}}. 
\end{eqnarray}
The Hamiltonian of the marginal interactions, 
\begin{eqnarray}\label{marg} 
H_{\rm{marg}} & = & \int dx [\lambda_{1} O_{1} + \lambda_{2} O_{2}], 
\nonumber \\ O_{1} & = & \psi^{1} \bar{\psi}^{1} \psi^{2} \bar{\psi}^{2} 
+ \psi^{1} \bar{\psi}^{1} \psi^{3} \bar{\psi}^{3} + \psi^{2} \bar{\psi}^{2} 
\psi^{3}\bar{\psi}^{3}, \nonumber \\ O_{2} & = & \psi^{0} \bar{\psi}^{0} 
(\psi^{1} \bar{\psi}^{1} + \psi^{2} \bar{\psi}^{2} + \psi^{3}\bar{\psi}^{3}), 
\end{eqnarray}
contains current-current contributions of the forms $({\bf J}_{1} + {\bf 
\bar{J}}_{1}) ({\bf \cdot J}_{2} + {\bf \bar{J}}_{2})$, arising from 
the interleg interactions, and $-({\bf J}_{1} {\bf \bar{J}}_{1} + {\bf J}_{2} 
{\bf \bar{J}}_{2})$ from the intraleg couplings, as well as a 
contribution from the normal-ordered marginal product $Tr(\sigma^{a}g) 
Tr(\sigma^{a}g') Tr(\sigma^{b}g) Tr(\sigma^{b}g')$ which originates 
in the four-spin term $({\bf\tilde{S}}_{i} {\bf \cdot \tilde{T}}_{i}) 
({\bf\tilde{S}}_{i+1}{\bf \cdot \tilde{T}}_{i+1})$. Explicit expressions 
for the marginal couplings $\lambda_{1}$ and $\lambda_{2}$ are given the 
Appendix.

\subsubsection*{Weak Coupling}

In the weak-coupling regime the results of the Appendix provide the 
expressions   
\begin{eqnarray}
m_{t} & = & J_{\perp} - 2J_{\times} - 20\lambda^{2} K, \nonumber\\
m_{s} & = & - 3J_{\perp} + 6J_{\times} + 12\lambda^{2} K,
\end{eqnarray}
for triplet and singlet masses, where $\lambda$ is another nonuniversal 
quantity. The marginal interactions renormalize these masses, and the 
phase transition occurs when the renormalized triplet mass vanishes, 
{\it i.e.} $m_{t}^{\rm ren} = 0$. On the other hand, the matrix-product 
ansatz (Sec.~III) gives the phase transition line exactly. One may 
therefore attempt to specify the renormalization by taking (for example 
at $J_{\times} = 0$) the value of $\lambda^{2}$ to be in agreement with 
the exact result, which suggests that $\lambda^{2} = 1/5$ for this 
second-order phase transition.

\subsubsection*{Strong Coupling}

In the strong-coupling regime we analyze perturbations around the CFT 
which corresponds to the exact solution of Eq.~(\ref{HB}). For this it 
is convenient to perform the transformation (\ref{dual}), which from 
Eq.~(\ref{ladham}) yields
\begin{eqnarray}
\!H & = & \sum_{i} {\textstyle \frac{1}{2}} K ({\bf\tilde{S}}_{i} {\bf 
\cdot \tilde{S}}_{i+1} \! + \! {\bf\tilde{T}}_{i} {\bf \cdot \tilde{T}}_{i+1})
+ (J_{\perp}\! + \! K)({\bf\tilde{S}}_{i} {\bf \cdot \tilde{T}}_{i}) 
\nonumber\\
& & \! + {\textstyle \frac{1}{2}} J_{\times} \{ {\bf \tilde{S}}_{i}
{\bf \cdot \tilde{T}}_{i+1} \! + \! {\bf \tilde{T}}_{i} {\bf \cdot 
\tilde{S}}_{i+1}
\! + \! {\bf\tilde{S}}_{i} {\bf \cdot \tilde{S}}_{i+1} \! 
+ \! {\bf \tilde{T}}_{i} {\bf \cdot \tilde{T}}_{i+1}\nonumber\\
& & \! - 4 [({\bf\tilde{S}}_{i} {\bf \cdot \tilde{S}}_{i+1}) 
({\bf\tilde{T}}_{i} {\bf \cdot \tilde{T}}_{i+1})
 - ({\bf\tilde{S}}_{i} {\bf \cdot \tilde{T}}_{i+1}) 
({\bf\tilde{S}}_{i+1} {\bf \cdot \tilde{T}}_{i})]\} \nonumber\\
& & \! + {\textstyle \frac{1}{2}} J_{||} \{ {\bf\tilde{S}}_{i} 
{\bf \cdot \tilde{S}}_{i+1} \! + \! {\bf\tilde{T}}_{i} {\bf \cdot 
\tilde{T}}_{i+1} 
\! + \! {\bf\tilde{S}}_{i} {\bf \cdot \tilde{T}}_{i+1} \! 
+ \! {\bf\tilde{T}}_{i} {\bf \cdot \tilde{S}}_{i+1}\nonumber\\
& & \! + 4 [({\bf\tilde{S}}_{i} {\bf \cdot \tilde{S}}_{i+1}) 
({\bf\tilde{T}}_{i} {\bf \cdot \tilde{T}}_{i+1}) 
-({\bf\tilde{S}}_{i} {\bf \cdot \tilde{T}}_{i+1})
({\bf\tilde{S}}_{i+1} {\bf \cdot \tilde{T}}_{i})]\}\nonumber\\
& & \! + 2K \sum_{i}({\bf\tilde{S}}_{i} 
{\bf \cdot \tilde{T}}_{i})({\bf\tilde{S}}_{i+1} {\bf \cdot 
\tilde{T}}_{i+1}).\nonumber
\end{eqnarray}
By substituting
\begin{eqnarray}
J_{L} & = & K/2, \label{eebbm} \nonumber\\
J_{D} & = & J_{\times}/2 + J_{||}/2, \nonumber\\
V_{LL} & = & - 2J_{\times} + 2J_{||}, \nonumber\\
V_{DD} & = & 2J_{\times} - 2J_{||},\\
V_{RR} & = & 2 K, \nonumber\\
J_{R} & = & J_{\perp} + K, \nonumber 
\end{eqnarray}
in (\ref{coupl}) the expressions for triplet and singlet masses become
\begin{eqnarray}
m_{t} & = & J_{\perp} + K - J_{\times} - J_{||} - \lambda^{2} (16 J_{||} 
+ 4K - 16 J_{\times}),\nonumber\\
m_{s} & = & - 3J_{\perp} - 3K + 3J_{\times} + 3J_{||} + 12\lambda^{2} K. 
\end{eqnarray}
The marginal current-current interactions act again to renormalize these 
masses. Setting $J_{\times} = 0$ and taking $J_{\perp} = J_{||} = J$ yields 
the value of the second-order phase transition point as $K/J = 
16 \lambda^{2} / (1 - 4\lambda^{2})$. A comparison with the DMRG results of 
Ref.~\onlinecite{Lah}, which suggest that at strong coupling the transition 
point is $K/J\approx 5-6$ [in the units of Eq.~(\ref{ladham})], yields 
$\lambda^{2} \approx 1/7$, a value rather close to that obtained in the 
weak-coupling regime. After this the equation $m_{t}=0$ gives the form 
of the critical surface in the strong-coupling limit, which is represented 
by the shaded triangle in the phase diagram of Fig.~\ref{PD} and by the 
bold, dashed line in Fig.~\ref{section}; the shaded triangle lies below 
the $Z_4$-symmetric plane, but does 
not include the origin of the coordinate systems. We note here that the 
perturbative considerations applied above are in a strict sense 
questionable, because in the problem under consideration the 
``perturbative'' rung-rung coupling is of the same order as the coupling 
corresponding to the unperturbed CFT, which is $K/2$. For a ladder system 
in which  $V_{RR}$ is not fixed to be $2K$, and is small in comparison 
with $V_{LL}$ and $V_{DD}$, the  second-order phase transition with 
central charge $c=3/2$ in the strong-coupling limit is accessible by  
perturbative analysis around the exact solution (\ref{HB}). Thus for 
consistency this term should be added to the Hamiltonian with a small 
coupling constant, $V_{RR}$, in order to study the RG behavior.     
We remark also that for $J_{\times}=J_{||}\equiv J$, the leg-leg and 
diagonal-diagonal biquadratic terms in Eq.~(\ref{eebbm}) cancel explicitly 
and the effective theory is described by massive fermions with a 
singlet-triplet mass splitting generated by $V_{RR}$ and $J_{R}$. 

The lines of phase transitions correspond to the vanishing of the triplet 
mass $m_{t}$. The theory is then equivalent to three massless fermions 
and is therefore described by a $c=3/2$ WZW model. Because of the $SU(2)$ 
symmetry of the ladder model (\ref{ladham}), this is a $SU(2)$ WZW model 
at level $k=2$. The bare singlet and triplet masses acquire a 
renormalization due to marginal interactions which can in principle 
be computed from the RG analysis. By considering the operator product 
expansion (OPE) for operators $O_{1}, O_{2}, \psi_{0}\bar{\psi}_{0}$, 
and $\psi_{a} \bar{\psi}_{a}$ ($a = 1,2,3$) (see Appendix and 
Eqs.~(\ref{bosspinvar},\ref{marg})), we deduce the one-loop RG equations 
\begin{eqnarray}
\frac{d\lambda_{1}}{d (\ln L)} & = & 2\pi(\lambda_{1}^{2}+\lambda_{2}^{2}), 
\ \frac{d\lambda_{2}}{d (\ln L)} = 4\pi\lambda_{1}\lambda_{2}, \nonumber\\
\frac{d g_{s}}{d (\ln L)} & = & g_{s}+3\pi g_{t}\lambda_{2}, \\
\frac{d g_{t}}{d (\ln L)} & = & g_{t}(1+2\pi\lambda_{1})+\pi g_{s}\lambda_{2}
\nonumber
\end{eqnarray}
for the marginal couplings $\lambda_{1}$ and $\lambda_{2}$, and for the 
singlet and triplet couplings $g_{s}$ and $g_{t}$, which are respectively 
proportional to $-m_{s}$ and $m_{t}$. A similar analysis is performed in 
Refs.~\onlinecite{NT} and \onlinecite{Sen}). The first and the second 
equations are decoupled from the others, and can be integrated (in 
variables $\lambda_{\pm} = {\textstyle \frac{1}{2}} (\lambda_{1} \pm 
\lambda_{2})$) 
to yield\cite{Sen}
\begin{eqnarray}
\lambda_{\pm} = \frac{\lambda_{\pm}^{(0)}}{1 - 8\pi\lambda_{\pm}^{(0)} 
\ln (L/L_{0})}.
\end{eqnarray}
From this it follows that for negative initial $\lambda_{\pm}$ these 
couplings are  marginally irrelevant (and renormalize to 0), whereas for 
positive values they are  marginally relevant (and develop exponential gaps). 
The initial (ultraviolet) fixed point is given by $g_{s}=g_{t}=0$, which 
corresponds to the CFT with $c=2$ (two uncoupled chains), and the RG 
equations for $g_{s}$ and $g_{t}$  yield the renormalization of masses 
under the RG flow. The new unstable fixed point is then defined from the 
renormalized value of $m_{t}$ as $m_{t}^{*}=0$. Because of these 
renormalization effects the exact transition line is difficult to 
estimate, but the presence of this transition is to be expected from 
$Z_{4}$ symmetry arguments.  

\subsubsection*{Self-dual model}

From the analysis above one may conclude that the lines (or surfaces) of the 
two second-order phase transitions with $c=3/2$ are not $Z_{4}$-symmetric, 
because the rung-rung coupling is proportional to $K$. The Hamiltonian which 
would have perfect symmetry of the two second-order phase transitions is 
\begin{eqnarray}
H & = & J_{||}\sum_{i}({\bf S}_{i} {\bf \cdot S}_{i+1} + {\bf T}_{i} 
{\bf \cdot T}_{i+1}) + J_{R}\sum_{i}({\bf\tilde{S}}_{i} 
{\bf \cdot \tilde{T}}_{i}), \nonumber \\
& & + {\textstyle \frac{1}{2}} K \sum_{i}({\bf\tilde{S}}_{i} {\bf \cdot 
\tilde{S}}_{i+1} + {\bf\tilde{T}}_{i} {\bf \cdot \tilde{T}}_{i+1}) \\  
& & + V_{RR}\sum_{i}({\bf\tilde{S}}_{i} {\bf \cdot \tilde{T}}_{i})
({\bf\tilde{S}}_{i+1} {\bf \cdot \tilde{T}}_{i+1}) \nonumber 
\end{eqnarray}
which corresponds to the model with $J_R$ and $V_{RR}$ independent of $K$. 
In the small-$J_{||}$ limit we have obtained two $SU(2)$ WZW models 
perturbed by relevant and marginal interactions which have their origin 
in the $K$ term and in the small couplings $J_{R}$ and $V_{RR}$. In the 
strong-$K$ regime there are again two $SU(2)$ WZW models perturbed by 
the small $J_{||}$, $J_{R}$, and $V_{RR}$ terms. We note that this 
Hamiltonian may be considered as a two-chain Hamiltonian $\sum_{i} 
P^{||}_{i} - E_{i}^-$ perturbed by terms proportional to 
$P_{i}^{||} - P_{i}^{\times}$, or as a $Z_{4}$-transformed Hamiltonian
with the $Z_{4}$-transformed perturbation. Qualitatively, the resulting 
behavior is the same in both cases. In the continuum limit this perturbation 
is represented by two four-fermion terms [see Eq.~(\ref{TLcontlim})], one 
of which gives the marginal contribution while the other gives the relevant 
contribution.  

\subsubsection*{Intermediate coupling}

 From the Zamolodchikov $c$-theorem,\cite{Zamol} a CFT 
perturbed by relevant and marginally relevant interactions will flow 
either to another CFT with smaller central charge or to a massive phase. 
For small $K$ and $J_{\perp}$, perturbation around the limit of decoupled 
chains generates a flow to a $c = 3/2$ CFT which is in the universality 
class of the bilinear-biquadratic $S = 1$ spin chain, a result established 
recently in Refs.~\onlinecite{rmvm} and~\onlinecite{rhn}. In the 
strong-coupling limit (large $K$) one may consider the $J_{||}$ term as a 
small perturbation and the same arguments are applicable (but now for the 
variables ${\bf\tilde{S}}$ and ${\bf\tilde{T}}$), the corresponding RG flow 
being from a critical $c = 2$ CFT to a $c = 3/2$ CFT. The universality class 
of this transition is the same as above: it is described by an 
$SU(2)$-symmetric CFT with level-$k = 2$ Kac-Moody algebra.   

From this one may conclude that the weak-coupling regime is not related 
continuously to the strong-coupling regime, and that there is a crossover 
between the two. The natural candidate for this crossover region 
would be the $Z_{4}$-symmetric plane, $J_{||} - J_{\times} = K$, probably 
in the vicinity of the point $J_{||} = J_{\perp} = K$ on the line 
$J_{\times} = 0$. However, it is important to address the question of 
whether this crossover is a phase transition or a continuous change. 
As shown in Ref.~\onlinecite{AH}, the $k = 2$ $SU(2)$ WZW model is 
unstable in the sense that it contains relevant operators which induce 
a flow to a stable $k = 1$ $SU(2)$ WZW, and one would therefore expect 
another second-order phase transition in the universality class of the 
$c = 1$ WZW model. The alternatives to this scenario are a first-order 
phase transition or a continuous crossover; a definitive statement is 
not possible on the basis of the present considerations alone. 

Because this special point or line cannot be expected to be accessible by 
RG analysis from either weak- or strong-coupling limits, it is natural 
to try to reach these points as a result of the flow away from the $c=3$ 
critical region defined in Sec.~IIB.3 [Eq.~(\ref{w})]. We thus consider 
the effect of relevant and marginal perturbations on the corresponding 
CFT originating from the different interactions in the Hamiltonian 
(\ref{ladham}).

\subsection{\mbox{\boldmath $SU(4)$}-symmetric basis}

It is clear that the most convenient way to obtain the general continuum 
limit is by direct fermionization of the $X$-operators which are the 
generators of the $SU(4)$ algebra. Before this we perform a canonical 
transformation of the $X_{i}^{10}$ and  $X_{i}^{30}$ operators for all 
lattice sites $i$,
\begin{eqnarray}
\tilde{X}^{10}_{i} & = & {\textstyle \frac{1}{\sqrt{2}}} i (X^{10}+i X^{30}) 
\longrightarrow X^{10}_{i},\nonumber\\
\tilde{X}^{30}_{i} & = & {\textstyle \frac{1}{\sqrt{2}}} i (X^{10}-i X^{30}) 
\longrightarrow X^{30}_{i},
\end{eqnarray}
from which one obtains [{\it cf.} Eqs.~(\ref{llpo},\ref{TLoper})]
\begin{eqnarray}\label{PAB} 
P_{i}^{||} & = & \sum_{\alpha\beta}X_{i}^{\alpha\beta}X_{i+1}^{\beta\alpha},
\nonumber \\
A^{\dag}_{i} & = & \sum_{\alpha} X^{\alpha 0}_{i}X^{\alpha 0}_{i+1}, \ 
\ B^{\dag}_{i}=\sum_{a}X^{00}_{i}X^{00}_{i+1}-X^{a0}_{i}X^{a0}_{i+1},
\nonumber \\
E_{i}^{+} & = & A_{i}^{\dag}A_{i} , \ \ E^{-}_{i}= B_{i}^{\dag}B_{i},
\end{eqnarray}
where $\alpha ,\beta = 0,\dots,3$ and $a = 1,2,3$.

The generators of the $su(4)$ algebra may be represented by four Dirac 
fermions,
\begin{eqnarray}\label{ferm}
X^{\alpha\beta}_{k}= {\textstyle \frac{1}{2}} (c_{k,\beta}c^{\dag}_{k,\alpha} 
- c^{\dag}_{k,\alpha} c_{k,\beta}),
\end{eqnarray}
subject to the local constraint $\sum_{\alpha =0}^{3}c^{\dag}_{k,\alpha} 
c_{k,\alpha}=1$.
In terms of these variables the $Z_{4}$ transformation \ref{Z4} is simply 
a particular case of the $U(1)$ canonical transformation for $c_{k,0}$ 
fermions,
\begin{eqnarray}
c_{k,0}\longrightarrow i c_{k,0}, \ \  c_{k,0}^{\dag}\longrightarrow
 - i c_{k,0}^{\dag}.
\end{eqnarray}
The fermionized version of the Hamiltonian (\ref{algham}) then takes the form
\begin{eqnarray}\label{cham}
\! H & = & \sum_{i,a,b}g_{1}(c_{i,a}^{\dag} 
c_{i,b}c_{i+1,b}^{\dag}c_{i+1,a}-\gamma c_{i,a}^{\dag}c_{i,b}c_{i+1,a}^{\dag} 
c_{i+1,b})\nonumber\\
& &\!\!+\,g_{2}(c_{i,a}^{\dag}c_{i,0}c_{i+1,0}^{\dag}c_{i+1,a} 
+ c_{i,0}^{\dag}c_{i,a}c_{i+1,a}^{\dag}c_{i+1,0})\nonumber\\
& &\!\!+\,g_{3}(c_{i,a}^{\dag}c_{i,0}c_{i+1,a}^{\dag}c_{i+1,0} 
+ c_{i,0}^{\dag}c_{i,a}c_{i+1,0}^{\dag}c_{i+1,a})\nonumber\\
& &\!\!+\,g_{4}(c_{i,a}^{\dag}c_{i,a}+c_{i+1,a}^{\dag}c_{i+1,a})\\
& &\!\!+\,g_{5}(1-c_{i,a}^{\dag}c_{i,a})(1-c_{i+1,a}^{\dag}c_{i+1,a}),\nonumber
\end{eqnarray}
where $g_{1} ={\textstyle \frac{1}{2}}( J_{||} + J_{\times} + K)$, $g_{2} = 
{\textstyle\frac{1}{2}}(J_{||} - J_{\times} + K)$, $g_{3} = {\textstyle 
\frac{1}{2}}(J_{||} - J_{\times} - K)$, $g_{4} = (J_{\perp} + 2K)$, $g_{5} 
= 2K$, and $\gamma = 1$. Note that we have retained the variable $\gamma$ 
as an additional perturbative parameter.

In the low-energy limit the lattice fermions are expressed in terms of 
left- and right-moving fermions $\psi_{L}, \psi_{R}$ at the Fermi points, 
\begin{eqnarray}
{\textstyle \frac{1}{\sqrt{a}}} c_{n,\alpha} \simeq \psi_{L,\alpha} 
\exp{(-ik_{F}x)} + \psi_{R,\alpha}\exp{(ik_{F}x)},
\end{eqnarray}
where $x = na$, $\pm k_{F} = \pm\pi/ra$, $a$ is the lattice constant, and 
$r$ the filling factor, which is related to the group dimension (2 for 
$SU(2)$, 4 for $SU(4)$). In the continuum limit the different operators 
may be classified according to their scaling dimensions in the vicinity 
of the conformally invariant points.

\subsubsection*{Perturbation around CFT with $c=3$}

For the particular set of values of the interaction parameters $J_{||}/K = 1$, 
$J_{\times}/K = 0$, which corresponds to a point in the $Z_{4}$-symmetric 
plane, the Hamiltonian may be written as [Eqs.~(\ref{PAB}), (\ref{algham})]
\begin{eqnarray}\label{sdual}
H&=& K\sum_{i}[P_{i}^{||}-\frac{1}{2}(E^{+}_{i}+E^{-}_{i})\nonumber\\
& &+2X_{i}^{00}X_{i+1}^{00}]-(J_{\perp}+2K)\sum_{i}X_{i}^{00}.
\end{eqnarray}
From the fermionic representation (\ref{ferm}) one observes that 
Eq.~(\ref{sdual}) corresponds to the  Hamiltonian (\ref{cham}) with the 
specific coupling values $g_{1} = g_{2} = g_{5}/2 = K$, $g_{3} = 0$, 
$\gamma = 1$ and $g_{4} = J_{\perp} + 2K$.  

In Sec.~IIB.3 we have discussed the soluble model which has $c = 3$ 
critical behavior in the region with finite rung and rung-rung interactions. 
Because this model is equivalent to the Hamiltonian (\ref{sdual}) when the 
second term vanishes identically ($\gamma = 0$), it would appear natural 
that the critical region be extended by an increase of the positive 
rung-rung interaction. We therefore consider the perturbation of this 
critical model caused by the term $(E^{+}_{i} + E^{-}_{i})$. In the 
continuum limit it is clear that the $\gamma$ term breaks the $SU(4)$ 
symmetry,
\begin{eqnarray}\label{TLcontlim}
c_{i,a}^{\dag} c_{i,b} c_{i+1,a}^{\dag} c_{i+1,b} & \! \simeq \! & 
\psi_{L,a}^{\dag} \psi_{L,b} \psi_{R,a}^{\dag} \psi_{R,b} + \dots,
\end{eqnarray}
producing a marginally relevant product of currents of the $SU(4)$-symmetric 
WZW model for $\gamma > 0$. The RG analysis\cite{IK} shows directly that this 
interaction opens an exponential gap and drives the system into a dimerized 
state. A similar effect occurs in the spin-orbital model with 
symmetry-breaking perturbations.\cite{AGLN,IQA} We note also that in the 
study of Ref.~\onlinecite{AffTL} this term appears as the continuum limit 
of the generators of the Temperley-Lieb algebra.\cite{cr} In the present 
analysis the combination ${\textstyle \frac{1}{2}} (E_{i}^{+} + E_{i}^{-}) 
- X_{i}^{00} X_{i+1}^{00}$ is the generator of the Temperley-Lieb algebra 
in the projected 3-state-per-site subspace (further details are provided in 
Ref.~\onlinecite{GB}). 

We have shown that the $SU(4)$-symmetric model (\ref{cham}) has an 
exponential gap, and this is expected to persist up to $\gamma \approx 1$,
at which point the model becomes that of Eq.~(\ref{ladham}) with parameters 
$J_{||}/K = 1$, $J_{\times}/K = 0$. Thus in the ring-exchange model 
(\ref{ladham}) not only this point is gapped but also the region around 
it in the $J_{\perp}$-direction. We note that the rung interaction 
[$g_{4}$ in Eq.~(\ref{cham})] induces both relevant (scaling dimension 3
/2) and marginal perturbations away from criticality.  

One may develop a perturbative analysis for the entire problem by starting 
from this $SU(4)$-symmetric solution. The $SU(4)_{k=1}$ CFT allows a 
decomposition into the direct sum of two $SU(2)_{k=2}$ CFTs, a fact 
used in the analysis of the $SU(4)$-symmetric spin-orbital model in 
Ref.~\onlinecite{IQA}. The addition of different perturbations to the 
$SU(4)$-symmetric solutions then leads to a lowering of the symmetry, 
and may induce a flow to one of the $SU(2)_{k=2}$ components. These 
correspond to the two second-order phase transitions with $c = 3/2$ at 
weak and strong coupling. The spin density is expressed in terms of the 
$SU(4)$ primary fields and therefore, in addition to the uniform ($k = 0$) 
and staggered ($k = \pi$) parts, has an oscillating part with momentum 
$k = \pi/2$. This result appears plausible for the problem at hand in 
that a semiclassical analysis\cite{Chub} reveals the presence of a ground 
state with orthogonal spin alignment on  neighboring sites. We will not 
dwell further on this approach here.   

\begin{figure}[t!]
\begin{center}
\includegraphics[width=8cm]{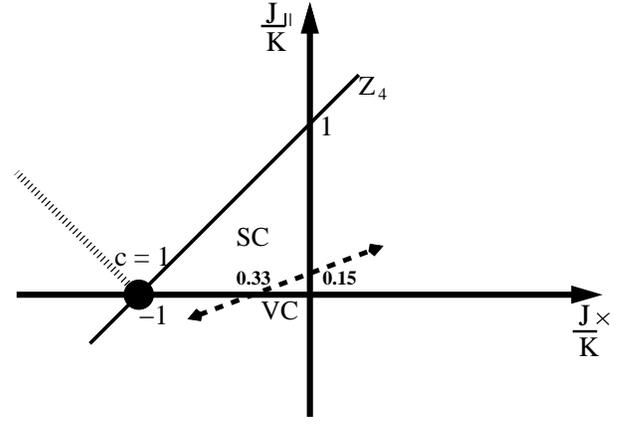}
\caption{Cross-section of the phase diagram (Fig.~\ref{PD}) in the plane 
$J_{\perp} = 0$. The solid line marks the projection of the plane of $Z_4$ 
symmetry, and the solid circle the line {\bf AB} of massless, incommensurate 
solutions. The dashed line represents the second Babujian-Takhtajan 
transition in the strong-coupling regime, which separates the scalar- and 
vector-chirality phases. The half-plane in which a massless region exists 
around the point $J_{\times} = - K$ is represented by the dotted 
line.\label{section}}
\end{center}
\end{figure}

\subsection{\mbox{\boldmath $c = 1$} CFT}

We conclude this section by considering  perturbations around the $c = 1$ 
CFT solution given by Eq.~(\ref{1/2}). Although the point $J_{\perp} = 0$ 
and line ${\bf AB}$ on the phase diagram of Fig.~\ref{PD} possess additional 
degeneracies, the conformal dimensions of the model are unaltered. The WZW 
model with $c = 1$ contains one primary matrix field of dimension 
$({\textstyle \frac{1}{4}},{\textstyle \frac{1}{4}})$, and the perturbation 
caused by this field is related to dimerization or alternation, {\it i.e.} 
it breaks explicitly the symmetry of translation by one lattice site. 
This type of interaction is absent in the initial Hamiltonian 
(\ref{ladham}), and therefore the relevant  perturbations due to this 
matrix field are disallowed by symmetry. The only possible perturbation 
is then a marginal current-current interaction which is present in the 
$g_{3}$ term of Eq.~(\ref{cham}). Perturbation around the exact solution 
given by Eq.~(\ref{1/2}) requires that all other couplings be set to zero. 
This corresponds to the ray $J_{||} + J_{\times} + K =0$ and $J_{\perp} = 0$ 
in the parameter space of the model, {\it i.e.} to the fermionized Hamiltonian
\begin{eqnarray}
H & = & {\textstyle \frac{1}{2}}\sum_{i,a}
[(c_{i,a}^{\dag} c_{i,0} c_{i+1,0}^{\dag} c_{i+1,a} + c_{i,0}^{\dag} 
c_{i,a} c_{i+1,a}^{\dag} c_{i+1,0}) \nonumber \\
& & + \, (1 - 2 c_{i,a}^{\dag}c_{i,a})(1 - 2 c_{i+1,a}^{\dag}c_{i+1,a})] \\
& & + \, g(c_{i,a}^{\dag} c_{i,0} c_{i+1,a}^{\dag} c_{i+1,0} + c_{i,0}^{\dag} 
c_{i,a} c_{i+1,0}^{\dag} c_{i+1,a}), \nonumber
\end{eqnarray}
where $g = J_{||} - J_{\times} - K$, and with the condition $J_{||} + 
J_{\times} + K = 0$, which follows from Eq.~(\ref{cham}). The general 
analysis of Ref.~\onlinecite{IK} is now applicable: when $g \equiv g_{3} 
< 0$ the symmetry-lowering term is marginally relevant and generates a 
gap, while if $g_{3} > 0$ the term is marginally irrelevant and one 
expects a massless region to extend along this ray, as represented 
in Fig.~\ref{section}. The point $g_{3} = 0$ represents a type of 
Berezinsky-Kosterlitz-Thouless transition, as shown already from the 
exact solution in Sec.~II. From the ray equation $J_{||} + J_{\times} 
+ K = 0$, the coupling $g$ is proportional to $J_{||}$, and thus 
for $J_{||} > 0$ one expects massless behavior, while for $J_{||} < 0$ 
the dynamical generation of a Haldane gap occurs. One may further expect 
that the massless behavior persists for a small region of values of 
$J_{\perp}$ around this ray in the plane $J_{||} + J_{\times} + K = 0$. 
For values of the parameters such that the phases develop a gap, 
incommensurate behavior is expected, in the form of a maximum in the 
structure function $S(q)$ at values $q = q_{\rm max}$ where $0 < q_{\rm max} 
< \pi$, over a broad region in the vicinity of the line {\bf AB} in 
Fig.~\ref{PD}.

\subsection{Universality classes}

The preceding subsections have revealed the rich critical behaviour of the 
system under investigation. There exist two critical surfaces described by 
CFTs with $c = 3/2$, the weak- and strong-coupling regimes, and one critical 
region in the universality class of a $c = 1$ CFT. The existence of further 
second-order phase transitions may be excluded on the basis of general 
arguments. Both the weak-coupling and the corresponding strong-coupling 
regions of the phase diagram may be obtained by considering perturbations 
around the fixed points of two decoupled spin chains. This corresponds to 
the flow from $c = 2$ CFTs to $c = 3/2$ CFTs: the $c = 3/2$ theories 
represents unstable fixed points, and a small change of the couplings in 
the Hamiltonian leads to further flow to the stable $c = 1$ CFT. For the 
general model (\ref{ladham}) with $SU(2)$ symmetry, the latter CFT has the 
lowest possible central charge. These considerations are consistent with 
the Zamolodchikov theorem. As discussed in Sec.~IVB, perturbations around 
the $SU(4)$-symmetric point reduce the symmetry and open a gap, thus 
excluding the possibility of $c = 3$ criticality. A candidate for this 
symmetry breaking is $SU(4) \rightarrow SU(3)$; however, the exactly soluble 
critical model with $c = 2$, which has $SU(3)$ symmetry,\cite{GB} is ``too 
far'' from the parameter space of the current model, in the sense that it 
has a positive diagonal-diagonal coupling $V_{DD}$, and addition of 
interaction terms to this model induces a flow to a massive phase. We 
therefore conclude that the list of second-order phase transitions 
presented in this section is complete.                 

\section{Summary and discussion}

We now collect all of the information presented in the preceding sections. 
The phase diagram is shown in Fig.~\ref{PD} in 3D form with coordinate 
axes $J_{||}/K$, $J_{\times}/K$, and $J_{\perp}/K$. We have chosen this 
convention to highlight the phases and phase transitions obtained at strong 
$K$, as relatively little is known about this regime. The parallel dot-dashed 
lines demarcate the $Z_{4}$-symmetric plane, which intersects the horizontal 
plane on the line $J_{\times}/K = -1$. This line lies parallel to the 
coordinate axis $J_{\perp}/K$, and the part $-4 < J_{\perp}/K < 4$, 
represented by the thick dashed line between points ${\bf A}$ and ${\bf B}$, 
forms a critical incommensurate region emerging from the exact solution 
obtained in Sec.~IIB and discussed in Sec.~III. 

The matrix-product ansatz (Sec.~III) reveals two lines of phase transitions, 
marked by the straight, solid lines in the $Z_{4}$-symmetric plane at 
$J_{\perp}/K = \pm 4$. The line {\bf AA'} at $J_{\perp}/K = 4$ represents 
a continuum of second-order phase transitions from a rung-singlet phase 
(Fig.~\ref{frs}) to a staggered dimer phase (Fig.~\ref{fsd}), the existence 
of which has been confirmed by a number of studies.\cite{rhn,dual,Lah} In 
Fig.~\ref{PD} we have shown only this line, which is known exactly, although 
in fact it constitutes a part of a surface. The line {\bf BB'} at $J_{\perp} 
/ K = - 4$ denotes a continuum of first-order phase transitions from a 
ferromagnetic phase (Fig.~\ref{ffm}) to a form of AKLT state 
(Fig.~\ref{faklt}). Both lines intersect the incommensurate line at their 
ends, which thus represent two multicritical points (${\bf A}$ and ${\bf B}$). 
We have deduced (Sec.~IV) the presence of another line (surface) of 
second-order phase transitions, the presence of which is consistent with 
the $Z_{4}$ symmetry of the plane in which it is in principle known exactly, 
and the conjectured form of the surface is represented both by the shaded 
triangle in the region close to the origin of Fig.~\ref{PD} and by the 
dashed line in Fig.~\ref{section}. This phase transition separates the 
strong-coupling vector-chirality phase (Fig.~\ref{fvc}), dual to the 
rung-singlet phase, from the scalar-chirality phase (Fig.~\ref{fsc}), 
which is dual to the staggered dimer phase. While the letters indicating 
the locations of the phases in Fig.~\ref{PD} refer in a strict sense only 
to the plane of exact $Z_4$ symmetry, the absence of phase transitions 
other than those present on the diagram means that the gapped phases 
obtained for model parameters outside the plane are connected continuously 
to the exactly known states, and their physical properties evolve 
continuously as the values of the interactions are further changed away 
from the $Z_4$ plane.

Renormalization-group techniques and perturbative approaches applied around 
the exactly known solutions may be used to determine the fixed-point structure 
of the entire phase diagram. The $Z_{4}$-symmetric plane is found also to 
constitute the transition region between weak- and strong-coupling regimes, 
which is centered on the star in the phase diagram of Fig.~\ref{PD}. Analysis 
of the relevance of the additional terms in the model perturbing the 
$c=3$ CFT  (Sec.~IV) suggest that all points on this 
plane, other than the transition lines of the preceding paragraph and the 
incommensurate line {\bf AB}, have gapped excitations. These spin gaps are 
generated only by marginal symmetry-breaking perturbations and are therefore 
small, a result which was confused with gapless behavior in the numerical 
studies of Ref.~\onlinecite{DMRG}. Within the plane, the RG fixed points 
corresponding to the $c = 3/2$ transitions are unstable (Sec.~V), and 
therefore we expect a flow to a stable $c = 1$ theory, for which the obvious 
candidate is the incommensurate line bounded by the multicritical points 
${\bf A}$ and ${\bf B}$. 

There is a certain mathematical and physical similarity between the 
behavior of the $S$ = 1/2 ladder with a ring-exchange interaction and 
the $S$ = 1 bilinear-biquadratic model described by the Hamiltonian
\begin{eqnarray}\label{BB}
H_{BB}=\sum_{i}({\bf S}_{i} {\bf \cdot S}_{i+1}) + \alpha ({\bf S}_{i} 
{\bf \cdot S}_{i+1})^{2}.
\end{eqnarray}
For different values of the parameter $\alpha$ this model contains a 
variety of phases and different forms of critical behavior. For $\alpha 
= -1$, the Babujian-Takhtajan model,\cite{BT} it is in the same universality 
class as the two $c = 3/2$ second-order phase-transition lines of the 
ring-exchange model. The Hamiltonian (\ref{BB}) also shows incommensurate 
behavior for the parameter region $1/3 < \alpha < 1$. However, in the 
bilinear-biquadratic model this is restricted to a gapped phase, in 
contrast to the current model where the critical incommensurate regime 
is related to the degeneracy of the $c = 1$ CFT. We comment that an 
explicit mapping from the $S$ = 1 model of Eq.~(\ref{BB}) to a 
$S$ = 1/2 ladder system may be obtained by using a composite-spin 
representation,\cite{rlfs} but that the resulting model does not lie 
in the phase space (Fig.~\ref{PD}) of a ladder with ring-exchange 
interactions (see Sec.~I). 

The most accurate and extensive numerical studies performed to date for a 
ladder system with ring exchange are contained in Ref.~\onlinecite{Lah}. The 
first quadrant of the circular phase diagram presented in this work 
corresponds to the line ($J_{||} = J_{\perp} > 0$, $J_{\times} = 0$) in 
Fig.~\ref{PD}. We note the complete agreement of our analysis with the 
numerical investigation for this line: the four phases of 
Ref.~\onlinecite{Lah} are those represented schematically in Figs.~\ref{frs}, 
\ref{fsd}, \ref{fsc}, and \ref{fvc}, and related under the $Z_4$ 
transformation as shown in Fig.~\ref{duality}. The complete phase diagram 
allows us not only to verify the Babujian-Takhtajan nature of the 
rung-singlet to staggered dimer phase transition but to confirm that the 
scalar-chirality to vector-chirality phase transition lies in the same 
universality class, and in addition to specify the location ($J_{||} = K$) 
and nature (Sec.~IV) of the crossover between staggered dimer and scalar 
chirality phases, which remained unclear from the numerical analysis. The 
second quadrant of Fig.~1 of Ref.~\onlinecite{Lah} is represented by the 
line ($J_{||} = J_{\perp} < 0$, $J_{\times} = 0$) in Fig.~\ref{PD}. 
Although we have not obtained specific information concerning this 
region of the phase diagram, the properties of the regions to which it 
is connected continuously allow us to deduce that the vector-chirality 
phase should be separated from the ferromagnetic phase by first-order 
transitions to a form of AKLT state, which indeed exhibits the 
collinear-spin property found by DMRG. Because our considerations do 
not include negative values of $K$, we refrain from comment on the 
third and fourth quadrants of the circular phase diagram.

Finally, we summarize briefly the relevance of our results for the 
materials and higher-dimensional systems mentioned in Sec.~I. Experimentally 
determined values of $K$ for cuprate systems, including the ladder compound 
La$_6$Ca$_8$Cu$_{24}$O$_{41}$, suggest that the ring-exchange interaction 
may in fact be close to the value required to drive the rung-singlet phase 
to a staggered dimer state, and thus that staggered dimer correlations 
may be detectable. In two dimensions one expects that a larger value of $K$ 
would be required to find analogs of the more exotic ( i.e. chiral) phases 
of the model, and thus that these would most likely be detectable, if at all, 
as excitations. Films of $^3$He have been found to offer larger values of 
$K$, but these appear both in combination with other multiple-spin exchange 
processes and on a more complex lattice geometry which may lead to further 
topological possibilities for spin configurations. Our results suggest that 
this restricted-geometry system may provide a very rich spectrum of possible 
phases, but a considerably more specific analysis would be required. 

In conclusion, we have analyzed a general model for a $S$ = 1/2 ladder 
with ring-exchange interactions. By investigating the exactly soluble points 
within the parameter space we obtain a complete classification of the phases 
and phase transitions in this type of system. Although we have considered the 
minimal model possessing both cyclic four-spin interactions and non-trivial 
exact solutions, we find a rich variety of gapped and gapless phases, of 
first- and second order-transitions, and of commensurate and incommensurate 
excitations, all connected by a complex renormalization-group flow pattern. 
The full phase diagram provides significant additional insight into the 
types of phases and transitions arising in low-dimensional spin systems 
as a consequence of the cooperation and competition between nearest-neighbor 
antiferromagnetic exchange interactions and multiple-spin interactions of 
the ring-exchange type. 

\begin{acknowledgments}

We would like to thank F. C. Alcaraz for explanations concerning the model 
of Eq.~(\ref{1/2}). We are grateful to A. Ferraz, P. Horsch, G. I. Japaridze, 
A. L\"{a}uchli, and C. Lhuillier for helpful communications and discussions. 
This work was supported by the Swiss National Science Foundation through 
grant \# 20-68047.02.

\end{acknowledgments}

\appendix

\section{}

We summarize here the important formulas used in Sec.~IV, and describe the 
fermionization procedure for the general bilinear-biquadratic spin-1/2 
ladder in the limits of two decoupled chains.

The continuum limit of the $SU(2)$ Heisenberg model is described by an 
$SU(2)$ WZW model at level $k=1$. This model has one matrix  primary 
operator $g_{\alpha}$ ($\alpha=0,1,2,3,4$) of scaling dimension $(1/4,1/4)$. 
The right and left Kac-Moody currents are fields of dimension $(1,0)$ and 
$(0,1)$. The relationship between these operators and the spin-operator 
density is given in the continuum limit by Eq.~(\ref{bosspinvar}). 
  
A pair of level-1 $SU(2)$ WZW models may be represented in terms of 4 Ising 
fields. The operator content of these Ising models provides a set of 
elementary variables for constructing the continuum limit of the general 
ladder model. The critical Ising model is described by a $c=1/2$ CFT in 
the continuum limit. It contains holomorphic and antiholomorphic fields, 
respectively $\psi(z)$ and $\bar{\psi}(\bar{z})$, with conformal dimensions 
$({\textstyle \frac{1}{2}},0)$ and $(0,{\textstyle \frac{1}{2}})$. The 
energy operator $\epsilon(z,\bar{z})=i\psi(z)\bar{\psi}(\bar{z})$ has 
dimension $({\textstyle \frac{1}{2}},{\textstyle \frac{1}{2}})$, while 
the order field $\sigma(z,\bar{z})$ and disorder field $\mu(z,\bar{z})$, 
related to the order field by Kramers-Wannier duality, have the same 
dimension $({\textstyle \frac{1}{16}},{\textstyle \frac{1}{16}})$. The 
holomorphic and antiholomorphic components of energy-momentum tensor are 
$T(z) = -{\textstyle \frac{1}{2}} \psi \partial \psi$ and $\bar{T} 
(\bar{z}) = -{\textstyle \frac{1}{2}} \bar{\psi} \bar{\partial} \bar{\psi}$. 

The representation of the fundamental level-1 $SU(2) \times SU(2)$ WZW 
fields is given in terms of 4 Ising models by (see e.g. Ref.~\onlinecite{Sen})
\begin{eqnarray}
g_{0} & = & \sigma_{1}\sigma_{2}\sigma_{3}\sigma_{0}
+ \mu_{1}\mu_{2}\mu_{3}\mu_{0},\nonumber\\
g_{1} & = & \mu_{1}\sigma_{2}\sigma_{3}\mu_{0} 
- \sigma_{1}\mu_{2}\mu_{3}\sigma_{0},\nonumber\\
g_{2} & = & \sigma_{1}\mu_{2}\sigma_{3}\mu_{0} 
+ \mu_{1}\sigma_{2}\mu_{3}\sigma_{0},\nonumber\\
g_{3} & = & \sigma_{1}\sigma_{2}\mu_{3}\mu_{0}
- \mu_{1}\mu_{2}\sigma_{3}\sigma_{0},\nonumber\\
\\
g_{0}' & = & \ \ \sigma_{1}\sigma_{2}\sigma_{3}\sigma_{0}
- \mu_{1}\mu_{2}\mu_{3}\mu_{0},\nonumber\\
g_{1}' & = & -\mu_{1}\sigma_{2}\sigma_{3}\mu_{0} 
- \sigma_{1}\mu_{2}\mu_{3}\sigma_{0},\nonumber\\
g_{2}' & = & -\sigma_{1}\mu_{2}\sigma_{3}\mu_{0}
+ \mu_{1}\sigma_{2}\mu_{3}\sigma_{0},\nonumber\\
g_{3}' & = & -\sigma_{1}\sigma_{2}\mu_{3}\mu_{0}
- \mu_{1}\mu_{2}\sigma_{3}\sigma_{0},\nonumber
\end{eqnarray}
and the expressions for the $SU(2)\times SU(2)$ Kac-Moody currents are
\begin{eqnarray}
J_{1} & = & {\textstyle \frac{1}{2}} i (\psi_{1}\psi_{0} 
- \psi_{2}\psi_{3}),\nonumber\\
J_{2} & = & {\textstyle \frac{1}{2}} i (\psi_{2}\psi_{0}
- \psi_{3}\psi_{1}),\nonumber\\
J_{3} & = & {\textstyle \frac{1}{2}} i (\psi_{3}\psi_{0}
- \psi_{1}\psi_{2}),\nonumber\\
\\
J_{1}' & = & - {\textstyle \frac{1}{2}} i (\psi_{1}\psi_{0}
+ \psi_{2}\psi_{3}),\nonumber\\
J_{2}' & = & - {\textstyle \frac{1}{2}} i (\psi_{2}\psi_{0}
+ \psi_{3}\psi_{1}),\nonumber\\
J_{3}' & = & - {\textstyle \frac{1}{2}} i (\psi_{3}\psi_{0} 
+ \psi_{1}\psi_{2}),\nonumber
\end{eqnarray}
The fields $g_{i}$ ($i$=1,2,3) represent the staggered part of the 
spin-density operator, while the Kac-Moody currents correspond to its 
uniform part.   

The operator-product expansions between Ising-model fields are\cite{FMS}
\begin{eqnarray}
\sigma(z,\bar{z})\sigma(w,\bar{w}) & \sim & \frac{1}{|z-w|^{\frac{1}{4}}}
+ {\textstyle \frac{1}{2}}|z-w|^{\frac{3}{4}} \epsilon(w,\bar{w}), \nonumber 
\\ \mu(z,\bar{z})\mu(w,\bar{w}) & \sim & \frac{1}{|z-w|^{\frac{1}{4}}} 
- {\textstyle \frac{1}{2}}|z-w|^{\frac{3}{4}}\epsilon(w,\bar{w}), \\
\! \sigma(z,\bar{z}) \mu(w,\bar{w}) & \! \sim & \! 
\frac{\gamma(z-w)^{\frac{1}{2}} \psi(w) + \gamma^{*} 
(\bar{z} - \bar{w})^{\frac{1}{2}} \bar{\psi}(\bar{w})}
{\sqrt{2}|z-w|^{\frac{1}{4}}}, \nonumber \\
\!\mu(z,\bar{z})\sigma(w,\bar{w}) \! & \! \sim \! & \! 
\frac{\gamma^{*}(z-w)^{\frac{1}{2}}\psi(w) 
+ \gamma(\bar{z}-\bar{w})^{\frac{1}{2}} \bar{\psi}(\bar{w})}
{\sqrt{2}|z-w|^{\frac{1}{4}}},\nonumber
\end{eqnarray}
and 
\begin{eqnarray}
\psi(z)\psi(w) & \sim & \frac{1}{z-w} + 2(z-w)T(w),\nonumber\\
\bar{\psi}(z)\bar{\psi}(w) & \sim & \frac{1}{\bar{z}-\bar{w}}
+ 2(\bar{z}-\bar{w})\bar{T}(\bar{w}), \nonumber \\
\psi(z) \sigma(w,\bar{w}) & \sim & \frac{\gamma\mu(w,\bar{w})}
{\sqrt{2}(z-w)^{\frac{1}{2}}},\nonumber\\
\psi(z)\mu(w,\bar{w}) & \sim & \frac{\gamma^{*}\sigma(w,\bar{w})}
{\sqrt{2}(z-w)^{\frac{1}{2}}},\\
\bar{\psi}(\bar{z}) \sigma(w,\bar{w}) & \sim & \frac{\gamma^{*} 
\mu(w,\bar{w})}{\sqrt{2}(\bar{z}-\bar{w})^{\frac{1}{2}}},\nonumber\\
\bar{\psi}(\bar{z})\mu(w,\bar{w}) & \sim & \frac{\gamma\sigma(w,\bar{w})}
{\sqrt{2}(\bar{z}-\bar{w})^{\frac{1}{2}}},\nonumber\\
\epsilon(z,\bar{z})\epsilon(w,\bar{w}) & \sim & \frac{1}{|z-w|^{2}},\nonumber
\end{eqnarray}
where $\gamma =\exp (i\pi/4)$.

These relations allow one to compute the OPEs between marginal operators 
defined in Eq.~(\ref{marg}), 
\begin{eqnarray} \label{OPE}
O_{1}(z)O_{1}(w)&\sim &\frac{3}{|z-w|^{4}}- \frac{2}{|z-w|^{2}}O_{1}+...,  
\nonumber\\
O_{2}(z)O_{2}(w) & \sim & \frac{3}{|z-w|^{4}}- \frac{2}{|z-w|^{2}}O_{1}+...,
\nonumber\\
O_{1}(z)O_{2}(w) & \sim & \frac{-2O_{2}}{|z-w|^{2}}+...,   \\
(\psi_{1}\bar{\psi}_{1})(z)O_{1}(w) & \sim & \frac{-(\psi_{2}
\bar{\psi}_{2})(w) - (\psi_{3}\bar{\psi}_{3})(w)}{|z-w|^{2}},\nonumber\\
(\psi_{2}\bar{\psi}_{2})(z)O_{1}(w) & \sim & \frac{-(\psi_{1}
\bar{\psi}_{1})(w) - (\psi_{3}\bar{\psi}_{3})(w)}{|z-w|^{2}},\nonumber\\
(\psi_{3}\bar{\psi}_{3})(z)O_{1}(w) & \sim & \frac{-(\psi_{2}
\bar{\psi}_{2})(w) - (\psi_{1}\bar{\psi}_{1})(w)}{|z-w|^{2}},\nonumber\\
(\psi_{a}\bar{\psi}_{a})(z)O_{2}(w) & \sim & \frac{-(\psi_{0}
\bar{\psi}_{0})(w)}{|z-w|^{2}} \ , \ (a=1,2,3),\nonumber\\
(\psi_{0}\bar{\psi}_{0})(z)O_{2}(w) & \sim & \nonumber\\
& &\!\!\!\!\!\!\!\!\!\!\!\!\frac{-(\psi_{1}\bar{\psi}_{1})(w)
- (\psi_{2}\bar{\psi}_{2})(w)-(\psi_{3}\bar{\psi}_{3})(w)}{|z-w|^{2}}.\nonumber
\end{eqnarray}

Because the scaling dimensions of all operators are known around 
CFT points, one may proceed to the continuum limit for the spin 
Hamiltonian around the limit of two decoupled chains described  by the 
WZW model with $c = 2$. There exist two variants of this limit, 
at weak and strong $K$, and it is convenient to study the continuum limit of 
the general Hamiltonian
\begin{eqnarray}\label{genham}
H & = & \sum_{i} J_{L}({\bf S}_{i} {\bf \cdot S}_{i+1} + {\bf T}_{i} 
{\bf \cdot T}_{i+1}) \nonumber \\
& & \; + J_{D}({\bf S}_{i} {\bf \cdot T}_{i+1} + {\bf T}_{i} 
{\bf \cdot S}_{i+1}) + J_{R}{\bf S}_{i} {\bf T}_{i} \nonumber \\
& & \; + V_{LL}({\bf S}_{i} {\bf \cdot S}_{i+1})({\bf T}_{i}
{\bf \cdot T}_{i+1}) \\
& & \; + V_{RR}({\bf S}_{i} {\bf \cdot T}_{i})({\bf S}_{i+1}
{\bf \cdot T}_{i+1}) \nonumber\\
& & \; + V_{DD}({\bf S}_{i} {\bf \cdot T}_{i+1})({\bf S}_{i+1} {\bf \cdot 
T}_{i})\nonumber
\end{eqnarray}
where in addition to the leg ($J_{L}$), rung ($J_{R}$), and diagonal 
($J_{D}$) Heisenberg interactions we include leg-leg ($V_{LL}$), 
rung-rung ($V_{RR}$) and diagonal-diagonal ($V_{DD}$) four-spin 
interactions. The most relevant contribution from the biquadratic 
terms arises from the product of the staggered parts of the corresponding 
composite quadratic expressions, and from the operator product expansion 
between quadratic product of currents with the quadratic products 
of the staggered parts,
\begin{eqnarray}
{\bf S}_{i} {\bf \cdot S}_{i+1} & \simeq & (O_{1}+O_{2})
+ (-1)^{x/a}\lambda g_{0},\nonumber\\
{\bf T}_{i} {\bf \cdot T}_{i+1} & \simeq & (O_{1}+O_{2})
+ (-1)^{x/a}\lambda g_{0}',\\
{\bf S}_{i} {\bf \cdot T}_{i} & \simeq & {\textstyle \frac{1}{2}}
(O_{1} - O_{2}) - (\epsilon_{1} + \epsilon_{2} + \epsilon_{3} 
- 3\epsilon_{0})\nonumber\\
& & \!\!\! + (-1)^{x/a}:[g^{a}(J^{a'} + \bar{J}^{a'}) 
+ (J^{a} + \bar{J}^{a})g'^{a}]:,\nonumber\\
{\bf S}_{i} {\bf \cdot T}_{i+1} & \simeq & {\textstyle \frac{1}{2}}
(O_{1} - O_{2}) + (\epsilon_{1} + \epsilon_{2} + \epsilon_{3}
- 3\epsilon_{0})\nonumber\\ 
& & \!\!\! + (-1)^{x/a}:[g^{a}(J^{a'} + \bar{J}^{a'}) 
- (J^{a} + \bar{J}^{a})g'^{a}]:,\nonumber
\end{eqnarray}
and
\begin{eqnarray}
\lefteqn{:({\bf S}_{i} {\bf \cdot S}_{i+1})::({\bf T}_{i}
{\bf \cdot T}_{i+1}):\simeq }\nonumber\\ 
& & \lambda^{2}[-4(O_{1}+O_{2})+6\sum_{k=0}^{3}\epsilon_{k}],\nonumber\\
\lefteqn{:({\bf S}_{i} {\bf \cdot T}_{i+1})::({\bf S}_{i+1}
{\bf \cdot T}_{i}):\simeq } \\
& & \lambda^{2} [O_{1} - 5O_{2} + 3\epsilon_{0} - 5(\epsilon_{1} 
+ \epsilon_{2}+\epsilon_{3})+3\sum_{k=0}^{3}\epsilon_{k}],\nonumber\\
\lefteqn{:({\bf S}_{i} {\bf \cdot T}_{i})::({\bf S}_{i+1} 
{\bf \cdot T}_{i+1}):\simeq } \nonumber \\
& & \lambda^{2}[O_{1} - 5O_{2} - 3\epsilon_{0} + 5(\epsilon_{1} 
+ \epsilon_{2} + \epsilon_{3}) - 3\sum_{k=0}^{3} \epsilon_{k}],\nonumber
\end{eqnarray}
where the cut-off-dependent constant $\lambda$ emerges from the operator 
product expansion between different contributions.

These expressions give in terms of Majorana fermions the continuum limit of 
the Hamiltonian, which is separated into singlet, triplet, and marginal parts 
as in Eq.~(\ref{contham}), with
\begin{eqnarray}\label{mass}
m_{t}&=& J_{R}-2J_{D}-\lambda^{2}(6V_{LL}-2V_{DD}+2V_{RR}),\label{coupl} 
\nonumber\\
m_{s}&=& -3J_{R}+6J_{D}-\lambda^{2}(6V_{LL}+6V_{DD}-6V_{RR}),\\
\lambda_{1}&=& -4J_{L}+J_{R}/2+J_{D}+(-4V_{LL}+V_{DD}+V_{RR}),\nonumber\\
\lambda_{2}&=& -4J_{L}-J_{R}/2-J_{D}-(-4V_{LL}-5V_{DD}-5V_{RR}),\nonumber
\end{eqnarray}
and operators $O_{1}$ and $O_{2}$ as defined in Eq.~(\ref{marg}). This 
representation allows one to study simultaneously the limits of weak and 
strong $K$ (Sec.~IV). We note that the signs of the four-spin interaction 
terms in the expression for the masses are different from those in 
Ref.~\onlinecite{rmvm} but the same as those resulting from the analysis 
in Ref.~\onlinecite{rfls} of the string order parameter.


\begin{thebibliography}{99}
\bibitem{cr}
The operators $P^{||}_{i,i+1}$, $P^{\times }_{i,i+1}$ and $E_{i,i+1}$, 
$\tilde{E}_{i,i+1}$ are generators of the algebra
\begin{eqnarray}
P_{i}^2 = 1,  \ &  & \   P_{i}P_{i+1}P_{i}=  P_{i+1}P_{i}P_{i+1}, \nonumber \\ 
E_{i}^2 = 4E_{i}, \ &  & \  E_{i}E_{i+1}E_{i}=  E_{i},   \nonumber \\  
P_{i}E_{i} = E_{i}P_{i}=E_{i}, & & \nonumber \\ 
P_{i}E_{i+1}E_{i}=  P_{i+1}E_{i}, \ & & \  E_{i+1}E_{i}P_{i+1} = 
E_{i+1}P_{i},\nonumber
\end{eqnarray}
for neighboring plaquettes, and
\begin{eqnarray}
P_{i}P_{j} = P_{j}P_{i},\  E_{i}E_{j} =  E_{j}E_{i},\  P_{i}E_{j} = E_{j}P_{i} 
\nonumber \end{eqnarray}
for $|i-j|>1$. We have used here the plaquette notation 
$P_{i,i+1}\equiv P_{i}$ and $E_{i,i+1} \equiv E_{i}$, in which $P_{i}$ 
denotes either $P^{||}_{i}$ or $P^{\times }_{i}$ and $E_{i}$ denotes 
either $E_{i}$ or $\tilde{E}_{i}$. We note that there exist a number of 
other relations between the operators which follow from the above. The 
operators $P_{i}$ satisfy the braid-algebra relations, while the operators 
$E_{i}$ form a Temperley-Lieb algebra.\cite{TL} Together these operators 
form a Brauer\cite{brauer} (or braid monoid\cite{AW}) algebra.
\end{thebibliography}
\end{document}